# ERRATUM: Energy dependence of the $\rho$ resonance in $\pi\pi$ elastic scattering from lattice QCD

Jozef J. Dudek,[1,2,*] Robert G. Edwards,[1] and Christopher E. Thomas[3]

(for the Hadron Spectrum Collaboration)

[1] *Thomas Jefferson National Accelerator Facility, Newport News, VA 23606, USA*
[2] *Department of Physics, Old Dominion University, Norfolk, VA 23529, USA*
[3] *DAMTP, University of Cambridge, Cambridge, UK*

Since the publication of [1], the authors have discovered a bug in the code which was used to perform the Wick contractions of quark fields in hadron correlation functions. The net effect of this bug is to cause an error in the normalization of those correlators with meson-meson constructions at both source and sink, $\langle \pi\pi(t)\,\pi\pi(0)\rangle$, relative to those which feature a single-meson operator at source and/or sink, $\langle \rho(t)\,\pi\pi(0)\rangle$, $\langle \pi\pi(t)\,\rho(0)\rangle$, $\langle \rho(t)\,\rho(0)\rangle$. The spectra obtained from variational analysis of these correlation matrices are in error, and it follows that the elastic scattering phase-shifts extracted using the finite-volume Lüscher formalism are also in error.

The authors have subsequently corrected the normalization and performed the variational analysis of the correct correlation matrices leading to new determinations of the finite-volume spectra, which show small but significant differences to those presented in [1]. A summary of the new spectra in the $\pi\pi$ elastic energy region is presented in Figure 1. The corresponding $P$-wave phase-shift points are presented in Figure 2, where they are observed to show a clear resonant behavior.

The conclusions of the paper do not change in qualitative content after the correction described above, with a resonant $\rho$ state still being present. A relativistic Breit-Wigner provides an excellent description of the $P$-wave phase-shift points in the elastic $\pi\pi$ scattering region, with parameter values, statistical uncertainties and correlations given by

$$a_t\, m_R = 0.15085(18)(3) \quad \begin{bmatrix} 1 & -0.26 \\ & 1 \end{bmatrix}$$
$$g = 5.80(10)(1)$$
$$\chi^2/N_{\mathrm{dof}} = \frac{24.5}{31-2} = 0.84\,.$$

This fit in physical units is presented in Figure 3.

The original, unedited manuscript, published as Ref. [1], follows this erratum in the pdf file.

[1] J. J. Dudek, R. G. Edwards and C. E. Thomas, Phys. Rev. D **87**, no. 3, 034505 (2013) [arXiv:1212.0830 [hep-ph]].

* dudek@jlab.org

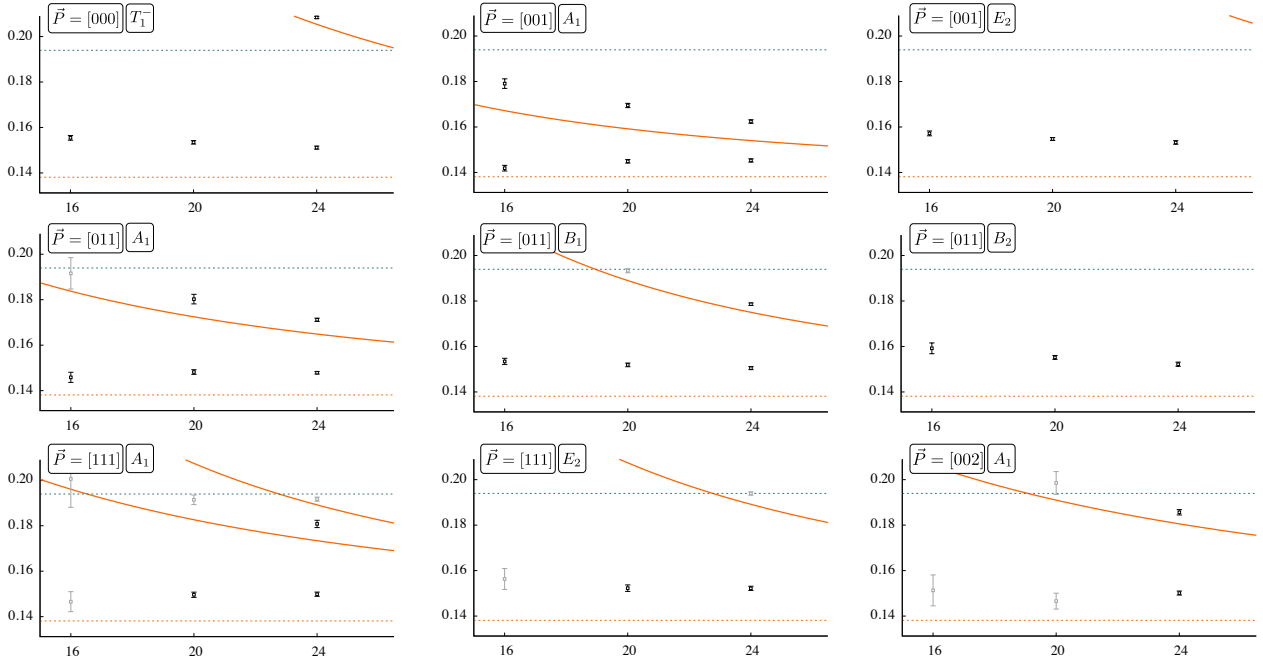

FIG. 1. Volume dependence of elastic spectra for various $\vec{P}$, $\Lambda$. Plotted is $a_t E_{\mathsf{cm}}$ versus $L/a_s$. Also shown by dashed horizontal lines are the $\pi\pi$ and $K\overline{K}$ energy thresholds. Solid curves indicate the non-interacting $\pi\pi$ energy levels. Points shown in gray are excluded from the phase-shift analysis.

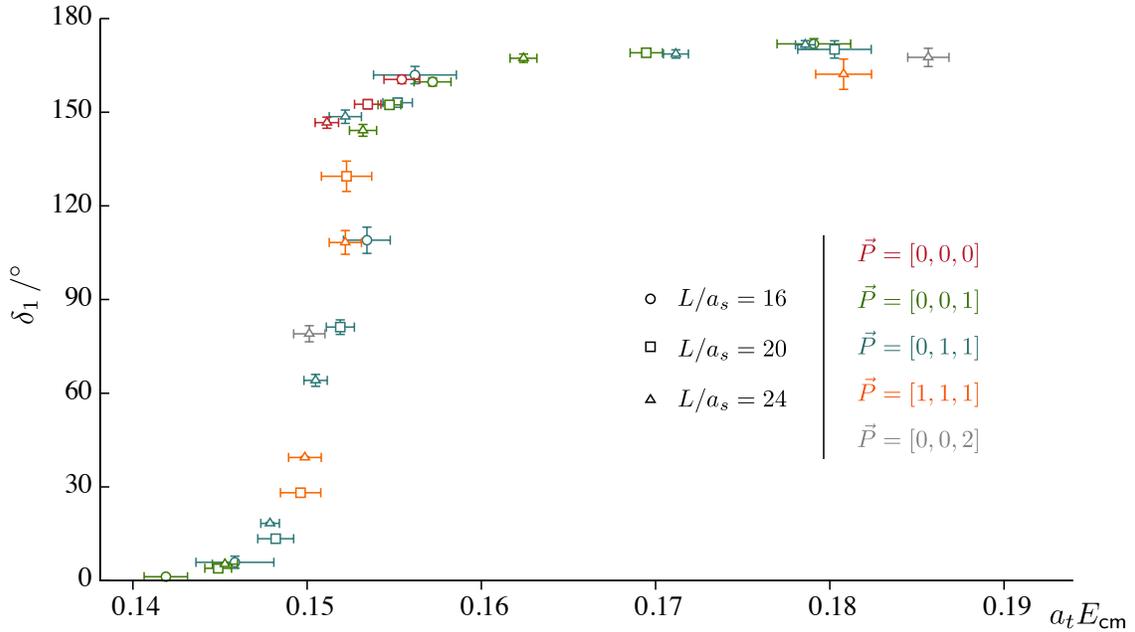

FIG. 2. $P$-wave $\pi\pi$ elastic scattering phase-shift, $\delta_1(E_{\mathsf{cm}})$, determined from solution of Eq. 7 in [1] applied to the finite-volume spectra shown in Fig. 1 under the assumption that $\delta_{\ell>1} = 0$. Energy region plotted is from $\pi\pi$ threshold to $K\overline{K}$ threshold.

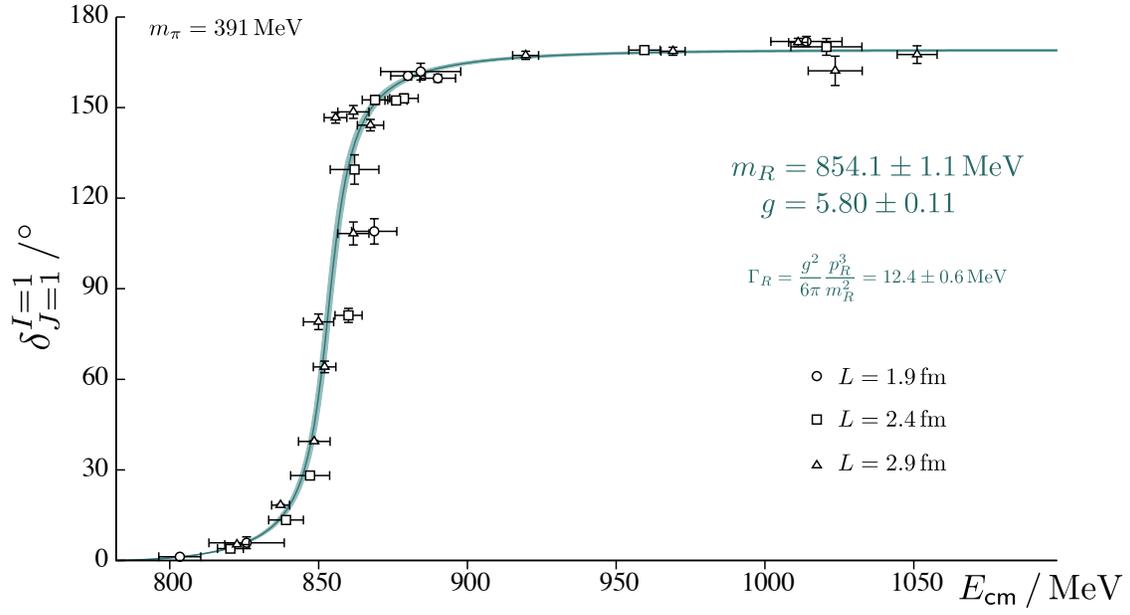

FIG. 3. Isospin-1, $P$-wave $\pi\pi$ elastic scattering phase shift and Breit-Wigner parameterisation for $m_\pi = 391\,\text{MeV}$. Energy region plotted is from $\pi\pi$ threshold to $K\overline{K}$ threshold.

# Energy dependence of the $\rho$ resonance in $\pi\pi$ elastic scattering from lattice QCD

Jozef J. Dudek,[1, 2, *] Robert G. Edwards,[1, †] and Christopher E. Thomas[3, ‡]

(for the Hadron Spectrum Collaboration)

[1] *Thomas Jefferson National Accelerator Facility, Newport News, VA 23606, USA*
[2] *Department of Physics, Old Dominion University, Norfolk, VA 23529, USA*
[3] *School of Mathematics, Trinity College, Dublin 2, Ireland*

We determine the energy-dependent amplitude for elastic $\pi\pi$ $P$-wave scattering in isospin-1 by computing part of the discrete energy spectrum of QCD in finite cubic boxes. We observe a rapidly rising phase shift that can be well described by a single $\rho$ resonance. The spectrum is obtained from hadron correlators computed using lattice QCD with light quark masses corresponding to $m_\pi \sim 400$ MeV. Variational analyses are performed with large bases of hadron interpolating fields including, as well as fermion bilinears that resemble $q\bar{q}$ constructions, also operators that resemble pairs of pions with definite relative and total momentum. We compute the spectrum for a range of center-of-mass momenta and in various irreducible representations of the relevant symmetry group. Hence we determine more than thirty values of the isospin-1 $P$-wave scattering phase shift in the elastic region, mapping out its energy dependence in unprecedented detail.

## I. INTRODUCTION

Hadron spectroscopy is principally the study of resonances which decay strongly, with widths of tens or hundreds of MeV, into asymptotic states corresponding to a multiplicity of hadrons stable under the strong interaction like the pion. Hadron resonances typically appear as enhancements in the continuous energy distribution of these multi-hadron final states. The simplest examples are elastic (or nearly elastic) resonances like the $\rho(770)$ which appears in the $I = 1$, $J = 1$ (isospin-1, spin-1) channel of $\pi\pi$ scattering, or the $\Delta(1232)$ in the $I = \frac{3}{2}$, $J = \frac{3}{2}$ channel of $\pi N$ scattering. In the elastic case the energy dependence of a partial-wave (definite-$J$) amplitude can be expressed in terms of a single real number, the phase-shift, which in the case of a narrow resonance will show a rapid rise from angles near $0°$, through $90°$, approaching $180°$ at energies above the resonance.

Although our only observables involve asymptotic many-hadron states, our desire is to understand hadron resonances at the level of interacting quarks and gluons within QCD, and this presents a significant theoretical challenge. The stable hadrons (e.g. pions) in the initial and final states are strongly interacting quark-gluon composites and the interactions that give rise to the resonant intermediate state contain nontrivial features like quark-antiquark annihilation. These complex and interrelated effects demand a consistent treatment of the nonperturbative dynamics of QCD.

Lattice QCD, in which the theory is formulated on a cubic Euclidean space-time grid of finite extent, has the advantage of being a first-principles approach to QCD in which the approximations required to render the theory computationally tractable are under control. It is a nonperturbative approach that, in the formulation we will use, respects unitarity - a particularly important feature since we plan to study resonances (like the $\rho$) which maximally saturate unitarity. The lattice grid spacing acts as a regulator for the field theory which can be progressively reduced toward the continuum. In practice the use of 'improved' discretised actions leads to discretisation errors which are small for many quantities of interest. The restriction to finite spatial volume turns out to be a tool that allows us indirect access to hadron scattering amplitudes. These amplitudes are not directly accessible via ($n > 2$)-point Euclidean correlation functions [1] but can be inferred using the discrete energy spectrum which follows from periodic boundary conditions applied to a finite volume. The formalism for relativistic elastic scattering in a cubic box is presented in [2, 3] for the case of a system in its rest-frame, with the extension to moving frames in [4–6].

A well known practical problem with the implementation of lattice QCD is the poor scaling of the required computation time for a realistic calculation with decreasing value of the quark mass. Only very recently have we seen calculations with quark masses low enough that the determined pion mass comes out near the experimental value, and they are typically of only the simplest of quantities (see a review in [7]). In this paper we will present calculations performed with a single strange quark of approximately the correct physical mass and two degenerate light quarks with a mass such that the pion has $m \sim 400$ MeV.

The challenge then is to compute the excited state spectrum of QCD in a finite-volume so that it can be related to scattering amplitudes. In a series of papers [8–14] we have explored the problem of extracting excited state energy spectra from large matrices of 2-point correlation functions, $\langle 0|\mathcal{O}_f(t)\mathcal{O}_i^\dagger(0)|0\rangle$, where $\mathcal{O}$ are composite op-erators with the quantum numbers of hadrons, built from quark and gluon fields. Using a large basis of such operators, we have extracted highly excited spectra using variational analysis. Computing with such large operator bases is made efficient using the *distillation* framework [15], which also renders simple the inclusion of quark annihilation diagrams with a high degree of statistical precision. This was demonstrated in the computation of the isoscalar meson spectrum in [13], where for example, the $\eta$ and $\eta'$ masses were determined with statistical uncertainty below 2% and even exotic isoscalar states with $J^{PC} = 1^{-+}$ above 2100 MeV could be cleanly extracted.

In [8, 9, 11, 13], the meson operator basis was limited to constructions of the type $\bar{\psi}\mathbf{\Gamma}\psi$, where $\mathbf{\Gamma}$ was constructed using Dirac gamma matrices and up to three gauge-covariant derivatives. The resulting spectra did not show the expected strong volume dependence of multi-hadron-like states, and the proposed explanation was that the fermion bilinear operators have only weak overlap onto such states. To remedy this we should augment the operator basis with some having large overlap onto multi-hadron-like states.

In [16, 17] we considered the case of the empirically non-resonant $I = 2$ $\pi\pi$ scattering using a basis of $\pi\pi$-like interpolating fields of the form $(\bar{\psi}\mathbf{\Gamma}_\pi\psi)_{\vec{p}_1} \cdot (\bar{\psi}\mathbf{\Gamma}_\pi\psi)_{\vec{p}_2}$, the use of *distillation* allowing for operators of definite relative momentum at both source and sink and thus enabling a variational analysis of matrices of correlation functions. By considering multiple frames in which the entire $\pi\pi$ system is in-flight and many irreducible representations of the reduced symmetry group of a boosted cube, the $S$ and $D$-wave scattering phase-shifts were mapped out across the elastic region $2m_\pi < E_{\rm cm} < 4m_\pi$, showing the expected weak repulsive interaction.

In this paper we will present results combining these two operator bases, forming matrices of correlation functions that include both constructions of the form $\bar{\psi}\mathbf{\Gamma}\psi$ and $(\bar{\psi}\mathbf{\Gamma}_\pi\psi)_{\vec{p}_1} \cdot (\bar{\psi}\mathbf{\Gamma}_\pi\psi)_{\vec{p}_2}$ with vector quantum numbers. The resulting variationally obtained spectrum leads to a detailed mapping of the $P$-wave phase shift for isospin-1 $\pi\pi$ scattering.

The $I = 1$, $J = 1$ scattering channel has been considered previously in lattice QCD calculations, [18–22]. For example, in [19], four quark masses ($m_\pi = 290 - 480\,{\rm MeV}$) were considered in two-flavor QCD calculations, with up to six points on the phase-shift curve determined at each pion mass. Fits to a simple Breit-Wigner form were used to suggest a quark mass dependence for the $\rho$ mass and width. In [20], a single pion mass, $m_\pi = 266\,{\rm MeV}$ in two-flavor QCD on a relatively small volume ($m_\pi L \sim 2.7$) was considered, with five points on the phase-shift curve determined; this was again fit with a simple Breit-Wigner form. One of our objectives in this paper is to map out the energy dependence through determination of many discrete values of the phase shift, conclusively demonstrating resonant behaviour and justifying a resonant parameterisation. This will be achieved through extraction of the spectrum on three volumes and for a large number of irreducible representations in moving frames[1].

| $(L/a_s)^3 \times (T/a_t)$ | $N_{\rm cfgs}$ | $N_{\rm t_{srcs}}$ | $N_{\rm vecs}$ |
|---|---|---|---|
| $16^3 \times 128$ | 479 | 4 – 8 | 64 |
| $20^3 \times 128$ | 603 | 4 | 128 |
| $24^3 \times 128$ | 553 | 2 – 6 | 162 |

TABLE I. The lattice ensembles and propagators used in this paper. The lattice sizes and number of configurations are listed, as well as the number of time-sources (which varies somewhat according to the correlator momentum and irrep) and the number of distillation vectors $N_{\rm vecs}$ (to be described in Section VI) featuring in the correlator construction.

In each irreducible representation, we will extract a spectrum of states from threshold up to high excitations with many states above the inelastic thresholds into $K\bar{K}, \pi\omega, \ldots$. Since we do not in this first study use interpolating fields that have good overlap onto these additional multi-hadron channels, we do not trust the determined spectrum outside the elastic region and restrict ourselves to the extraction of the elastic $P$-wave phase-shift between the $\pi\pi$ and $K\bar{K}$ thresholds.

The remainder of the paper is organised as follows. In Section II we review the details of our lattice setup and in Section III we describe the method by which excited spectra are extracted from correlation functions. Section IV highlights the importance of including multi-hadron-like operators, in Section V we discuss the consequences of performing calculations on a lattice in a finite volume which has a reduced symmetry, and in Section VI we summarise how correlators are constructed using the distillation technique. We show the extracted finite-volume spectra in Section VII, in Section VIII we present the energy-dependent phase shift which follows from these spectra, and we give a summary in Section IX.

## II. FINITE VOLUME LATTICE GAUGE FIELDS

In Euclidean time, excited state contributions to correlation functions decay faster than the ground state, and at large times they are swamped by the signals of lower states, thus complicating the extraction of excited states. To ameliorate this problem we have adopted a dynamical anisotropic lattice formulation of Clover fermions with two light quarks and one strange quark. In this anisotropic formulation, the temporal extent is discretised with a finer lattice spacing than the spatial directions [26, 27], allowing a more precise resolution of the discrete time-dependence of correlation functions. In this work, computations were performed with spatial lattice spacing $a_s \sim 0.12\,{\rm fm}$, and a temporal lattice spacing ap-proximately 3.5 times smaller, corresponding to a temporal scale $a_t^{-1} \sim 5.6$ GeV. Results are presented for quark mass parameters $a_t m_l = -0.0840$ and $a_t m_s = -0.0743$ corresponding to a pion mass of 391 MeV, and on lattice sizes of $16^3 \times 128$, $20^3 \times 128$ and $24^3 \times 128$ with corresponding spatial extents $L \sim 2$ fm, $\sim 2.5$ fm and $\sim 3$ fm. Some details of the lattices and propagators used for correlation constructions are provided in Table I. The masses of the lightest stable mesons containing light and strange quarks, as determined on these lattices, are shown in Table II.

---

[1] advocating the use of moving frames is currently a popular trend [23–25].

| | |
|---|---|
| $a_t m_\pi$ | 0.06906(13) |
| $a_t m_K$ | 0.09698(9) |
| $a_t m_\eta$ | 0.10406(56) |
| $a_t m_\omega$ | 0.15678(41) |
| $\xi$ | 3.444(6) |

TABLE II. Stable meson masses and the anisotropy $\xi = a_s/a_t$ determined on the lattice ensembles listed in Table I. Pion and kaon masses come from an infinite-volume extrapolation, while the $\eta$ and $\omega$ masses are those evaluated on the $24^3$ lattice.

As well as the volume dependence of energies of multi-hadron states originating in hadronic interactions which are the subject of this work, there can also be dependence of single-hadron energies on $L$. These polarization effects, corresponding to a single hadron 'sensing' itself around the periodic volume, are typically largest for the lightest hadron, the pion, and can be characterized by the product $m_\pi L$. At the single quark mass value used in this study, $m_\pi L$ ranges from 3.8 to 5.7, and since effects typically fall off exponentially with increasing $m_\pi L$ we expect them to be rather small here. As an example, previous investigations have found [16, 28] that the pion mass variation with $m_\pi L$ on these lattices is not large.

In an infinite volume of continuous, isotropic spacetime, a single free particle can have any value of momentum with magnitude varying continuously up from zero, with the energy related to the momentum by the relativistic dispersion relation $E_{\vec{p}}^2 = m^2 + |\vec{p}|^2$. In a finite $L \times L \times L$ volume with periodic spatial boundary conditions, the momentum of a free particle is restricted to the discrete set $\vec{p} = \frac{2\pi}{L}(n_x, n_y, n_z) = \frac{2\pi}{L}\vec{n}$, with the $n$'s being a triplet of integers. We will write momenta in units of $\frac{2\pi}{L}$ with square brackets, e.g. $\vec{p} = [n_x, n_y, n_z]$.

A further complication which arises from our use of an anisotropic lattice is the need to determine the precise value of the anisotropy, $\xi$, which relates the spatial lattice spacing $a_s$ to the temporal lattice spacing $a_t = a_s/\xi$. The anisotropy appears in the dispersion relation of a free-particle, where as found previously [17], the dispersion relation for the pion (and other stable hadrons) can be well described by a continuum-like form

$$(a_t E_{\vec{n}})^2 = (a_t m_\pi)^2 + \frac{1}{\xi^2} \left( \frac{2\pi}{L/a_s} \right)^2 |\vec{n}|^2, \qquad (1)$$

for a range of lattice volumes. The result $\xi = 3.444(6)$ is used throughout the rest of this paper.

## III. EXTRACTING EXCITED STATE SPECTRA

Our determination of the spectrum of eigenstates of QCD in a finite volume proceeds through the calculation of matrices of correlation functions between suitable hadronic creation and annihilation operators at time 0 and $t$ respectively,

$$C_{ij}(t) = \langle 0|\mathcal{O}_i(t)\mathcal{O}_j^\dagger(0)|0\rangle,$$

For each correlation matrix, the set of operators $\{\mathcal{O}_i\}$, constructed from color-singlet combinations of quark and gluon fields, all have the same conserved quantum numbers. Within the basis of operators used we can attempt to find the optimal linear combination for interpolation of each possible finite-volume eigenstate $|\mathfrak{n}\rangle$. A commonly-used method to achieve this is a variational solution [2, 29, 30] and herein our particular application of the variational method follows that developed in Refs. [8, 9, 31]. A system of generalized eigenvalue equations is established for the correlation matrix

$$C(t)v^{\mathfrak{n}}(t) = \lambda_{\mathfrak{n}}(t)C(t_0)v^{\mathfrak{n}}(t), \qquad (2)$$

where $\lambda_{\mathfrak{n}}$ and $v^{\mathfrak{n}}$ are eigenvalues and eigenvectors for a state labelled by $\mathfrak{n}$. Eq. (2) is solved for the eigenvalues, and the exponential dependence on the Euclidean time, $\lambda_{\mathfrak{n}}(t) \sim e^{-E_{\mathfrak{n}}(t-t_0)}$, is used to determine the energy $E_{\mathfrak{n}}$ of the state. The orthogonal[2] eigenvectors represent the optimal combination of the operators $\mathcal{O}_i$ for interpolation of the state $|\mathfrak{n}\rangle$ from the vacuum, $\left(\mathcal{O}_{\mathfrak{n}}^{\text{opt.}}\right)^\dagger \sim \sum_i v_i^{\mathfrak{n}} \mathcal{O}_i^\dagger$.

Any two-point correlation function on a finite spatial lattice can be expressed as a spectral decomposition

$$C_{ij}(t) = \sum_{\mathfrak{n}} \frac{Z_i^{\mathfrak{n}*} Z_j^{\mathfrak{n}}}{2E_{\mathfrak{n}}} e^{-E_{\mathfrak{n}} t}, \qquad (3)$$

where the "spectral overlap factors", $Z_i^{\mathfrak{n}} \equiv \langle \mathfrak{n}|\mathcal{O}_i^\dagger|0\rangle$ are related to the eigenvectors by $Z_i^{\mathfrak{n}} = \sqrt{2E_{\mathfrak{n}}} e^{E_{\mathfrak{n}} t_0/2} v_j^{\mathfrak{n}*} C_{ji}(t_0)$. The form of Eq. 3 assumes that $t \ll T$, the temporal length of the box, so that the contributions arising from other time orderings on the periodic lattice can be ignored. However, in practice, these alternate time-ordered contributions have a small but discernible effect on the determined energies; a method to account for them was presented in Ref. [17] and is used again in this work.

## IV. INCLUDING MULTI-HADRON OPERATORS

Although we have previously reported on the extraction of a significant number of excited meson states

---

[2] $v^{\mathfrak{n}\dagger} C(t_0) v^{\mathfrak{m}} = \delta_{\mathfrak{n},\mathfrak{m}}$

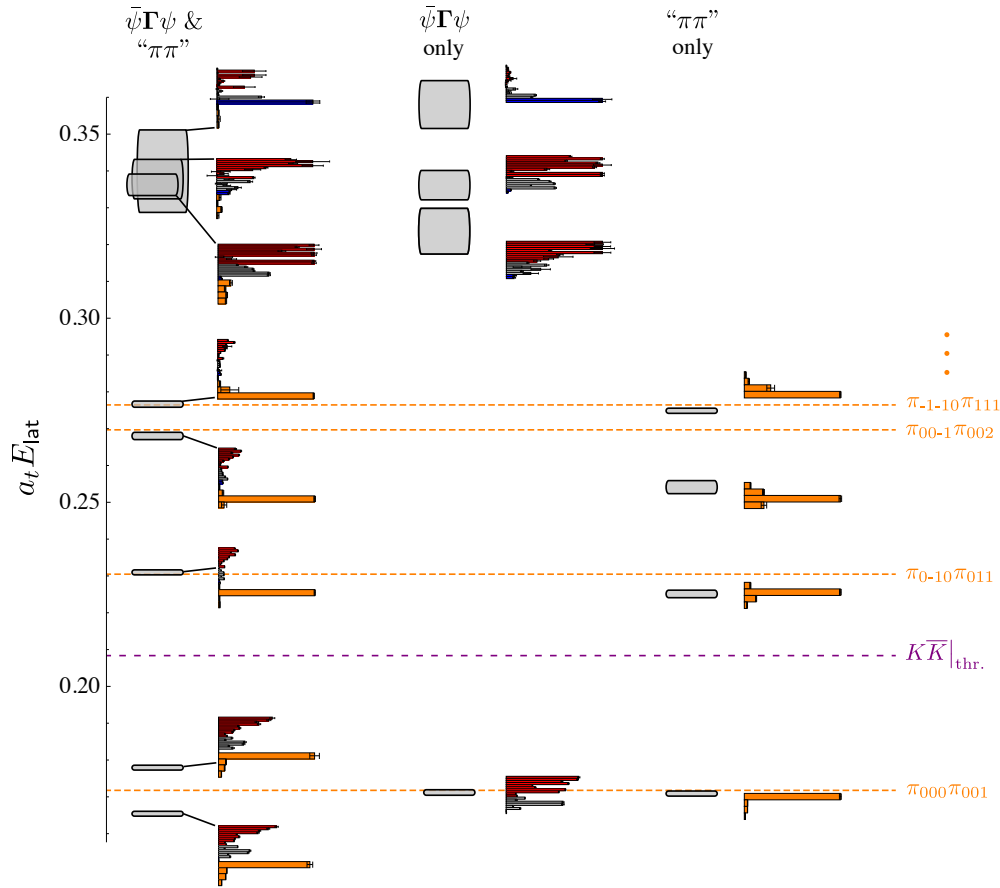

FIG. 1. The extracted spectrum of states, on the $24^3$ volume, using variational analysis on a matrix of correlation functions with total momentum $\vec{P} = [0,0,1]$ transforming irreducibly in the $A_1$ representation of the relevant group. The three columns left to right correspond to a large basis featuring operators of both fermion bilinear structure and $\pi\pi$ structure, just fermion bilinear and just $\pi\pi$ respectively. The histograms show the relative size of matrix element overlaps $\langle \mathfrak{n}|\mathcal{O}^\dagger|0\rangle$ for a range of operators. The red, gray and blue bars are fermion bilinears while the orange bars (drawn somewhat wider to be more visible) are $\pi\pi$ operators. Also shown by the dashed orange lines are the energies of non-interacting $\pi\pi$ pairs, $E_{\pi\pi}^{\text{n.i.}} = \sqrt{m_\pi^2 + \vec{n}_1^2 \left(\frac{2\pi}{L}\right)^2} + \sqrt{m_\pi^2 + \vec{n}_2^2 \left(\frac{2\pi}{L}\right)^2}$, and the dashed purple line shows the position of the $K\overline{K}$ threshold above which the system is inelastic.

[8, 9, 11, 13, 14], those calculations used only operators built from a single fermion bilinear at rest or in flight, $\mathcal{O}_i \sim \bar{\psi}\Gamma_i\psi$. A large basis was built using up to three gauge-covariant derivatives acting on the fermion fields, and this allowed many states to be extracted from variational analysis of correlator matrices. It was found that the extracted spectrum, which could be largely described in terms of constituent $q\bar{q}$ constructions supplemented with 'hybrid' states in which a gluonic excitation is also present, showed only very weak dependence on the lattice volume. In particular, the spectrum lacked signatures of multi-meson states, which if non-interacting would appear with a strongly volume-dependent spectrum and a characteristic distribution across irreducible representations of the lattice symmetry group. In the interacting theory we expect the actual eigenstates to be (volume-dependent) admixtures of what we might call "single-hadron" basis states and multi-hadron basis states.

Our explanation for this observation of a restricted spectrum is that the fermion bilinear operators have an overlap onto multi-hadron states that is suppressed by powers of the lattice volume, such that they couple only very weakly into the multi-hadron sector. To the extent that the eigenstates in any particular volume are admixtures of single-hadron and multi-hadron states, one can argue that correlators computed using fermion bilinears contain contributions from *all* eigenstates, $|\mathfrak{n}\rangle$, of the appropriate symmetry in $\sum_{\mathfrak{n}} \frac{Z_i^{\mathfrak{n}*} Z_j^{\mathfrak{n}}}{2E_{\mathfrak{n}}} e^{-E_{\mathfrak{n}} t}$. This is true, but there remains the practical problem of extracting the spectrum, which is particularly challenging when there are nearly-degenerate states, a likely occurrence in a dense spectrum of hadron scattering states. An effective solution, which does not depend upon distinguishing small energy differences, is to make use of orthogonality within the variational method, as described in the pre-vious section, where each state is optimally interpolated by a different orthogonal linear superposition of the basis operators even if they are approximately degenerate.

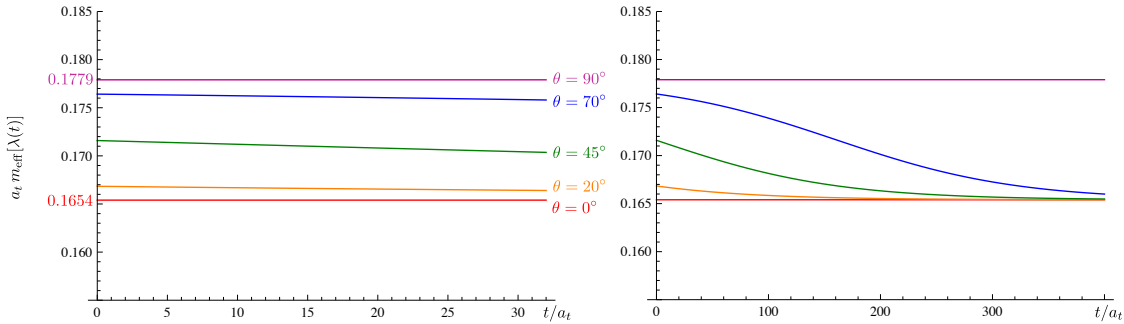

FIG. 2. Effective mass of Eq. 4 with $a_t E_1 = 0.1654$, $a_t E_2 = 0.1779$ for a range of mixing angles, $\theta$. Left panel shows typical time-separations in our calculation over which the time-dependence is essentially flat. Right panel indicates that eventually the presence of two low-lying states could be detected but only through observation of unrealistically long time separations with high statistical precision.

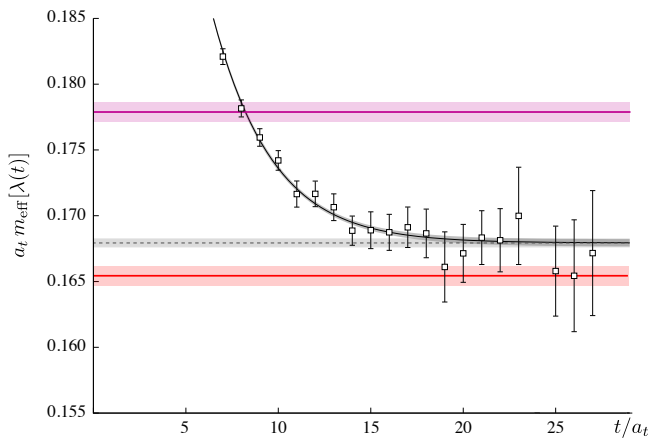

FIG. 3. Effective mass of lowest principal correlator obtained from variational analysis using only fermion bilinear operators (more time sources were used here than presented in the central column of Fig. 1). The gray band indicates a fit to the principal correlator using two exponentials as described in the text. The dashed gray line is the lower energy in that fit. Red and purple bands indicate the energies obtained from the lowest *two* principal correlators when using the full operator basis (with a smaller number of time sources).

The lack of operators that have significant overlap on multi-hadron states causes the variational analysis to be unable to form orthogonal eigenvectors in the larger space $\{|^{\text{single}}_{\text{hadron}}\rangle, |^{\text{multi}}_{\text{hadron}}\rangle\}$, since they have access only to $\{|^{\text{single}}_{\text{hadron}}\rangle\}$. We present a simple model below of what can happen if only a restricted operator basis is used when the true eigenstates of the system in finite-volume are ad-mixtures of single-hadron and multi-hadron basis states.

In [17] we demonstrated the construction of $\pi\pi$-like operators built from products of variationally optimised pion operators of definite momentum, $\pi^\dagger(\vec{k})$. In simple terms we construct

$$(\pi\pi)^{[\vec{k}_1, \vec{k}_2]\dagger}_{\vec{P},\Lambda} = \sum_{\vec{k}_1, \vec{k}_2} \mathcal{C}(\vec{P}, \Lambda; \vec{k}_1; \vec{k}_2)\, \pi^\dagger(\vec{k}_1)\, \pi^\dagger(\vec{k}_2)\ ,$$

for definite total momentum $\vec{P} = \vec{k}_1 + \vec{k}_2$ in irreducible representation, $\Lambda$. The sum is over the different directions, for fixed magnitude, of the definite pion momenta $\vec{k}_1, \vec{k}_2$, with the weights $\mathcal{C}$ ensuring that the operators transform in $\Lambda$. Through the use of multiple possible choices of magnitudes, $|\vec{k}_1|, |\vec{k}_2|$, we can build a basis of operators.

The importance of including $\pi\pi$-like operators in addition to fermion bilinears is demonstrated in Fig. 1 which shows the extracted spectrum computed on the $24^3$ lattice when the system has a total momentum of $\vec{P} = [0, 0, 1]$ and lies in the irreducible representation $A_1$, roughly corresponding to helicity-0 (to be discussed more in the next section). The three columns show the spectrum extracted using the full basis of operators (fermion bilinears plus $\pi\pi$-like), only fermion bilinear operators, and only $\pi\pi$-like operators. There are striking differences, notably the large gap between $a_t E_{\text{cm}} = 0.20$ and 0.30 which, when $\pi\pi$ operators are included, is populated by several states which have large overlap with the $\pi\pi$ operators. For our current purpose, the most important difference is the presence of two states in the elastic region with the full basis while only one appears with the restricted operator bases. Notice that in the full basis these two states each have large overlap onto *both* the fermion bilinear operators and the $\pi\pi$ operator of lowest relative momentum.

The essential problem with using a restricted operator basis can be demonstrated using a simple two-state mixing hypothesis, where the state basis is a "would-be" stable single hadron state, $|\rho\rangle$, and a multi-hadron state, $|\pi\pi\rangle$. The finite-volume energy eigenstates are linear superpositions of these two basis states:

$$|E_1\rangle = \cos\theta|\rho\rangle + \sin\theta|\pi\pi\rangle$$
$$|E_2\rangle = -\sin\theta|\rho\rangle + \cos\theta|\pi\pi\rangle.$$

A variational analysis that used multiple operators with good overlap onto both $|\rho\rangle$ and $|\pi\pi\rangle$ would be able to resolve the two orthogonal combinations, with two separate principal correlators having time dependences $\sim e^{-E_1 t}$, $\sim e^{-E_2 t}$. However, if one restricts to only operators with good overlap onto $|\rho\rangle$, the variational method will not be able to form two orthogonal vectors and the principal correlator will have a time-dependence similar to

$$\langle\rho|e^{-Ht}|\rho\rangle = \cos^2\theta\, e^{-E_1 t} + \sin^2\theta\, e^{-E_2 t}, \qquad (4)$$

which features both energy scales. In practice these energies can be very close together, e.g. $a_t E_1 = 0.1654(7)$ and $a_t E_2 = 0.1779(7)$ giving $a_t \Delta E = 0.013(6)$ in the case presented in Fig. 1. In Fig. 2 we show the effective mass[3] corresponding to Eq. 4 for a range of mixing angles, $\theta$. One clearly sees that over the time-separations typically resolved in lattice QCD calculations and, considering typical statistical uncertainty, the time-variation is compatible with a single-state hypotheses. It is only at very large times that the correlator relaxes to the true ground state and the correlator can be seen to clearly contain more than one state at a low energy scale. In practice correlators will also contain excited-state pollution from higher energy scales that impact at small Euclidean times.

In Fig. 3 we show the effective mass of the lowest principal correlator obtained from variational analysis using only the fermion bilinear operators. The gray curve shows the effective mass corresponding to a fit to the principal correlator, between $t/a_t = 5$ and $t/a_t = 41$, of the form $\lambda(t) = (1-A)e^{-E(t-t_0)} + Ae^{-E'(t-t_0)}$ where $a_t E = 0.16793(28)$ and $a_t E' = 0.494(11)$. The $\chi^2/N_{dof}$ for the fit is below 1.0 indicating that the data is described well in terms of one low-lying state and excited state contributions at a much higher energy scale. Clearly, from this principal correlator alone, we cannot infer the presence of two low-lying energy levels at 0.1654 and 0.1779.

In summary, if one wishes to determine the spectrum reliably, it proves necessary to include explicit multi-hadron operators into a variational basis. In this paper we will consider the inclusion of operators resembling a pair of pions of definite total and relative momentum. Since we do not yet include any operators resembling pairs of kaons, the next lightest stable hadrons, we will be cautious about trusting extracted levels near the top of the elastic region where they may be mixing with $K\overline{K}$ basis states.

In the next section we present in somewhat more detail the construction of two-meson operators relevant for a study of $\pi\pi$ scattering in isospin-1.

## V. MESON-MESON OPERATORS ON A FINITE CUBIC LATTICE

The symmetry of a lattice with a finite extent is reduced compared to that of an infinite volume continuum. In our implementation, we have a cubic lattice discretisation in a cubic box with periodic boundary conditions. The appropriate symmetry group is the double cover of the octahedral, or cubic, group with parity, $O_h^D$, and this is the symmetry relevant for a system of hadrons overall at rest. The allowed momenta are quantised by the boundary conditions giving $\vec{p} = \frac{2\pi}{L}(n, m, p)$, or $\vec{p} = [n, m, p]$ in our notation, where $L$ is the spatial extent of the lattice (in physical units) and $n, m, p$ are integers. For a system 'in flight', i.e. with a non-zero overall momentum, $\vec{P} \neq \vec{0}$, the symmetry is reduced further to that of the *little group* [32], which we denote by $LG(\vec{P})$, the subgroup of $O_h^D$ under which $\vec{P}$ is invariant.

The consequences of this reduced symmetry were discussed in detail in Ref. [17] and here we briefly review the salient points relevant for $\pi\pi$ scattering in isospin-1. For $\vec{P} = \vec{0}$, the continuum spin, $J$, is no longer a good quantum number on the lattice and states are instead labelled by the irreducible representations, *irreps*, of the octahedral group, of which there are a finite number. Parity, $P$, and any relevant flavour quantum numbers are still good. The manner in which the various components of a $J^P$ state are distributed, or *subduced*, into the irreps, $\Lambda^P$, is presented in Table II of Ref. [17].

The situation is more complicated for $\vec{P} \neq \vec{0}$; the pattern of subductions of the various *helicities*, $\lambda$, into the little group irreps depends on $LG(\vec{P})$, i.e. the type of momentum $\vec{P}$, see Ref. [11]. Table II of Ref. [17] shows these subductions. Note that, apart from the $\lambda = 0$ components, parity is *not* a good quantum number for a system in flight, but any relevant flavour quantum numbers are still good.

A state consisting of two identical hadrons must have a definite symmetry under the interchange of the two hadrons; it must be symmetric under this interchange if the hadrons are bosons or antisymmetric if they are fermions. This generalises to systems of two hadrons, having identical masses, related by a symmetry, e.g. $\pi^+$, $\pi^-$ and $\pi^0$ if isospin is a good symmetry, or $K$ and $\overline{K}$ related by charge-conjugation. In particular, taking into account the symmetry of the spin and any flavour parts of the wavefunction, the spatial wavefunction must have some definite symmetry under this interchange and so is restricted to either only even or only odd partial waves[4], $\ell$. For $\pi\pi$ in isospin-1, the flavour (isospin) wavefunction is antisymmetric and so Bose symmetry restricts the system to odd partial waves (with parity $P = (-1)^\ell = -1$ when $\vec{P} = \vec{0}$). The reduced symmetry of the finite volume lattice means that multiple partial waves appear in a single irrep as shown in Table III.

---

[3] $\mathrm{meff}\big[C(t)\big] = \frac{1}{\delta t}\log\frac{C(t)}{C(t+\delta t)}$

| $\vec{P}$ | LG($\vec{P}$) | $\Lambda^{(P)}$ | $\pi\pi$ $\ell^N$ |
|---|---|---|---|
| [0,0,0] | $O_h^D$ | $T_1^-$ | $1^1,\ 3^1$ |
| | | $T_2^-$ | $3^1$ |
| | | $A_2^-$ | $3^1$ |
| [0,0,n] | Dic$_4$ | $A_1$ | $1^1,\ 3^1$ |
| | | $E_2$ | $1^1,\ 3^2$ |
| | | $B_1$ | $3^1$ |
| | | $B_2$ | $3^1$ |
| [0,n,n] | Dic$_2$ | $A_1$ | $1^1,\ 3^2$ |
| | | $A_2$ | $3^1$ |
| | | $B_1$ | $1^1,\ 3^2$ |
| | | $B_2$ | $1^1,\ 3^2$ |
| [n,n,n] | Dic$_3$ | $A_1$ | $1^1,\ 3^2$ |
| | | $A_2$ | $3^1$ |
| | | $E_2$ | $1^1,\ 3^2$ |
| [n,m,0] | C$_4$ | $A_1$ | $1^2,\ 3^4$ |
| [n,n,m] | | $A_2$ | $1^1,\ 3^3$ |

TABLE III. The pattern of subductions of $I = 1$ $\pi\pi$ partial waves, $\ell \le 3$, into lattice irreps, $\Lambda$, where $N$ is the number of embeddings of this $\ell$ in this irrep. This table is derived from Table II of Ref. [17] by considering the subductions of the $\ell$ when $\vec{P} = \vec{0}$ and the subductions of the various helicity components for each $\ell$ when $\vec{P} \ne \vec{0}$. Here $\vec{P}$ is given in units of $\frac{2\pi}{L}$ and $n, m$ are non-zero integers with $n \ne m$. We show the double-cover groups but only give the irreps relevant for integer spin.

We note that, in contrast, for a two-meson state such as $\pi\pi'$, where $\pi'$ represents an excited pion, or $\pi\omega$, there is in general no such Bose symmetry constraint. For such a system overall at rest, parity, $P \propto (-1)^\ell$, is a good quantum number and so, even on a finite volume lattice, odd and even partial waves do not appear together in any irrep[5]. However, when $\vec{P} \ne \vec{0}$, parity is not a good quantum number, there is in general no definite symmetry under the interchange of the two hadrons[6] and all $\ell$ can mix; there is no separation between odd and even partial waves. In the current work, we restrict ourselves to scattering below inelastic thresholds and so do not consider

---

[4] e.g. if isospin is a good symmetry, even (odd) $\pi\pi$ partial waves have isospin $I = 0$ or 2 ($I = 1$) with positive (negative) G-parity. $K\overline{K}$ states with positive (negative) charge-conjugation parity, $C$, $\sim (K\overline{K} + C\ \overline{K}K)$, only occur in even (odd) partial waves

[5] e.g. non-interacting $\pi(\vec{k})\omega(-\vec{k})$ and $\pi(-\vec{k})\omega(\vec{k})$ have the same energy and appropriate linear combinations are eigenstates of parity

[6] e.g. non-interacting $\pi(\vec{k}_1)\omega(\vec{k}_2)$ and $\pi(\vec{k}_2)\omega(\vec{k}_1)$ with $\vec{P} = \vec{k}_1 + \vec{k}_2$ do not in general have the same energy

such mixing between odd and even partial waves. This will however be relevant as we go higher up in energy and in other scattering channels.

We use the single and multi-meson operators, respecting the symmetries of the finite volume lattice, which have been developed in a series of papers [8, 9, 11, 17]. For a single meson at rest, we first construct an operator with definite continuum $J^P$ and $J_z$-component $M$, $\mathcal{O}^{J^P,M}(\vec{k} = \vec{0})$, consisting of a fermion bilinear featuring gauge-covariant derivatives and Dirac gamma matrices, and with the correct flavour structure. We then subduce these into the relevant octahedral group irreps to form lattice operators [8, 9]. At non-zero momentum, our construction considers the subduction of an operator with definite continuum helicity, $\lambda$, $\mathbb{O}^{J^P,\lambda}(\vec{k})$, where $\vec{k}$ is quantised as discussed above, into the relevant little group irreps [11]. The general pattern of subductions is summarised in Table II of Ref. [17].

In Ref. [17] we discussed the single-meson operators relevant for a pion, which has $I = 1$ and negative $G$-parity, at rest and in flight. At rest, the subduction of a $J^P = 0^-$ operator into the irrep $\Lambda^P = A_1^-$ is trivial,

$$\mathcal{O}_{A_1^-}^{[0^-]}(\vec{0}) = \mathcal{O}^{0^-}(\vec{0})\ .$$

At non-zero momentum, for all the momenta we consider, the subduction into the $A_2$ irrep is also trivial,

$$\mathbb{O}_{A_2}^{[0^-]}(\vec{k}) = \mathbb{O}^{0^-}(\vec{k})\ .$$

When we use the variational method to find the optimal linear combination of operators to interpolate a pion, we will include operators subduced into $A_1^-$ from other $J$ (for $\vec{P} = \vec{0}$) and other helicity, $\lambda$, subduced into $A_2$ (for $\vec{P} \ne \vec{0}$). We use $\pi(\vec{k})$ as a shorthand to represent these optimal operators in the appropriate irreps.

For the $\rho$ meson, which has $I = 1$ and positive $G$-parity, at rest the three-dimensional $T_1^-$ irrep is a simple representation of $J^P = 1^-$ and the subduction is straightforward; this is given explicitly for our basis choice in Appendix A of Ref. [9]. In flight, the $\lambda = 0$ components subduce into the $A_1$ irrep and, depending on LG($\vec{P}$), the $\lambda = \pm 1$ components subduce into either the two-dimensional $E_2$ irrep or the one-dimensional $B_1$ and $B_2$ irreps [11].

As mentioned above and discussed in detail in Ref. [11], for a system in flight, parity is in general not a good quantum number. The helicity-0 components have a 'reflection parity', $\tilde{\eta} \equiv P(-1)^J$, where $J$ and $P$ are respectively the spin and parity at rest. For the momenta we consider, $\lambda = 0$ with $\tilde{\eta} = +(-)$ subduces into the $A_1$ ($A_2$) irrep and so there is no mixing between these even on a finite volume lattice. However, for $\lambda \ne 0$ there is no such symmetry and therefore, for example, the $\lambda = \pm 1$ components of the $\rho(1^-)$ appear in the same lattice irreps as the $\lambda = \pm 1$ components of the $b_1(1^+)$ (both these mesons have the same flavour structure). This is not relevant in the current work where we are restricting ourselves to relatively low energies, below inelastic thresholds, but must be taken into account if we wish to consider higher energies. Similar situations occur in other scattering channels.

In Section IV we have argued that to reliably determine the complete spectrum we need operators that have good overlap onto multi-hadron basis states. We find that operators built from the product of two or more single-hadron operators of the form described above are able to achieve this. We will follow the construction used in Ref. [17] where a general $\pi\pi$ creation operator is,

$$(\pi\pi)^{[\vec{k}_1,\vec{k}_2]\dagger}_{\vec{P},\Lambda,\mu} = \sum_{\substack{\vec{k}_1\in\{\vec{k}_1\}^\star \\ \vec{k}_2\in\{\vec{k}_2\}^\star \\ \vec{k}_1+\vec{k}_2=\vec{P}}} \mathcal{C}(\vec{P},\Lambda,\mu;\;\vec{k}_1;\vec{k}_2)\,\pi^\dagger(\vec{k}_1)\,\pi^\dagger(\vec{k}_2)\;,$$

(5)

where $\pi(\vec{k})$ is a single-pion operator and $\mathcal{C}$ is a Clebsch-Gordan coefficient for $\Lambda_1 \otimes \Lambda_2 \to \Lambda$ with $\Lambda_{1,2} = A_1^-$ of $O_h^D$ if $\vec{k}_{1,2} = \vec{0}$ and $\Lambda_{1,2} = A_2$ of $LG(\vec{k}_{1,2})$ if $\vec{k}_{1,2} \neq \vec{0}$, and where $\Lambda$ is an irrep of $LG(\vec{P})$. The sum over $\vec{k}_{1,2}$ is a sum over all momenta in the *stars* of $\vec{k}_{1,2}$, which we denote by $\{\vec{k}_{1,2}\}^\star$, and by which we mean all momenta related to $\vec{k}_{1,2}$ by an allowed lattice rotation. The $\left[\vec{k}_1,\vec{k}_2\right]$ label on the $\pi\pi$ operator indicates that it was constructed from single-pion operators with momenta in $\{\vec{k}_1\}^\star$ and $\{\vec{k}_2\}^\star$. We refer to Ref. [17] for further details, a discussion of the Clebsch-Gordan coefficients, and explicit values of $\mathcal{C}$.

As discussed in Ref. [17], we use "optimised" single-pion operators, $\pi(\vec{k})$, in order to obtain $\pi\pi$ correlators which are dominated by ground state pions at smaller times, i.e. to reduce contamination from $\pi\pi'$, where $\pi'$ is any higher mass state with pion quantum numbers. We follow that reference and construct variationally optimal operators by diagonalising a matrix of correlators in large basis of operators for each relevant irrep. As described in Section III, the $\mathfrak{n}^{\text{th}}$ eigenvector gives the variationally optimal linear combination of the basis operators to overlap with state $\mathfrak{n}$. Our basis consists of all single-meson operators subducing into the relevant irreps constructed from any possible Dirac gamma matrix structure and 0,1,2,3 derivatives at rest [9] or 0,1,2 derivatives in flight [11]. The efficacy of these operators was demonstrated in Ref. [17] and their use allows us to perform analyses at smaller Euclidean times.

In the current work we restrict ourselves to overall momenta $\vec{P} = [0,0,0],[0,0,1],[0,1,1],[1,1,1]$ and $[0,0,2]$. The various combinations of $\vec{k}_1$ and $\vec{k}_2$ used in our $\pi\pi$ operator constructions are presented in Table IV. For each lattice irrep, we include in our basis these $\pi\pi$ operators along with all relevant isospin-1 single-meson operators from any possible Dirac gamma matrix structure combined with 0,1,2,3 derivatives at rest [9] or 0,1,2 derivatives in flight [11]. The number of operators we use in each irrep is summarised in Table V.

| $\vec{P}$ | Volumes | $\vec{k}_1$ | $\vec{k}_2$ | $\Lambda^{(P)}$ |
|---|---|---|---|---|
| $[0,0,0]$ $O_h^D$ | $16^3,20^3,24^3$ | $[0,0,1]$ | $[0,0,-1]$ | $T_1^-$ |
| | | $[0,1,1]$ | $[0,-1,-1]$ | $T_1^-$ |
| | $16^3$ | $[1,1,1]$ | $[-1,-1,-1]$ | $T_1^-$ |
| $[0,0,1]$ $\text{Dic}_4$ | $16^3,20^3,24^3$ | $[0,0,0]$ | $[0,0,1]$ | $A_1$ |
| | | $[0,0,1]$ | $[0,0,0]$ | $A_1$ |
| | | $[0,-1,0]$ | $[0,1,1]$ | $A_1,E_2,B_1$ * |
| | | $[-1,-1,0]$ | $[1,1,1]$ | $A_1,E_2,B_2$ * |
| | | $[0,0,-1]$ | $[0,0,2]$ | $A_1$ |
| $[0,1,1]$ $\text{Dic}_2$ | $16^3,20^3,24^3$ | $[0,0,0]$ | $[0,1,1]$ | $A_1$ |
| | | $[0,1,0]$ | $[0,0,1]$ | $B_1$ |
| | | $[-1,0,0]$ | $[1,1,1]$ | $A_1,B_2$ |
| | | $[1,1,0]$ | $[-1,0,1]$ | $B_1,B_2$ |
| | | $[0,1,-1]$ | $[0,0,2]$ | $A_1,B_1$ |
| $[1,1,1]$ $\text{Dic}_3$ | $16^3,20^3,24^3$ | $[0,0,0]$ | $[1,1,1]$ | $A_1$ |
| | | $[1,0,0]$ | $[0,1,1]$ | $A_1,E_2$ |
| | | $[2,0,0]$ | $[-1,1,1]$ | $A_1,E_2$ |
| $[0,0,2]$ | $16^3,20^3,24^3$ | $[0,0,0]$ | $[0,0,2]$ | $A_1$ |

TABLE IV. The two-pion operators used presented for each $\vec{P}$ on various volumes; also shown is $LG(\vec{P})$. Example momenta $\vec{k}_1$ and $\vec{k}_2$ are shown; all momenta in $\{\vec{k}_1\}^\star$ and $\{\vec{k}_2\}^\star$ are summed over in Eq. 5. Swapping around $\vec{k}_1$ and $\vec{k}_2$ gives the same operators up to an overall phase. (*these $B_1$ and $B_2$ were not considered on the $16^3$ volume)

| $\vec{P}$ | Irrep | Single-meson | $\pi\pi$ $20^3,24^3$ $(16^3)$ |
|---|---|---|---|
| $[0,0,0]$ | $T_1^-$ | 26 | 2 (3) |
| $[0,0,1]$ | $A_1$ | 18 | 4 (4) |
| | $E_2$ | 29 | 2 (2) |
| | $B_1$ | 9 | 1 (0) |
| | $B_2$ | 9 | 1 (0) |
| $[0,1,1]$ | $A_1$ | 27 | 3 (3) |
| | $B_1$ | 29 | 3 (3) |
| | $B_2$ | 29 | 2 (2) |
| $[1,1,1]$ | $A_1$ | 21 | 3 (3) |
| | $E_2$ | 35 | 2 (2) |
| $[0,0,2]$ | $A_1$ | 18 | 1 (1) |

TABLE V. Number of single-meson and $\pi\pi$ operators used for each $\vec{P}$ and irrep on the various volumes. The two-meson operators are listed in Table IV and all relevant single-meson operator structures are considered including up to 3 derivatives at rest and up to 2 derivatives at non-zero momentum.

## VI. CORRELATOR CONSTRUCTION THROUGH DISTILLATION

In Section IV we have emphasised the need to include operators that have a strong overlap with expected multi-meson states and the desire to perform a variational analysis of a matrix of correlation functions forces us to find a correlator construction method that allows for such operators at both source and sink. In the isospin-1 channel, the contractions to form these correlators will involve quark-line 'annihilation' – that is diagrams which feature propagation of a quark from $t$ to the same $t$. In the previous section we described how multi-hadron operators transforming irreducibly under the relevant symmetry can be constructed from products of single-hadron operators of definite momentum. The projection into definite relative momentum requires a sampling of all spatial sites on a timeslice. The correlator construction method known as *distillation* [15] satisfies all the above desiderata in a natural and efficient way.

Distillation is a quark-field smearing method which is designed to increase overlap onto the low modes relevant for low-lying hadronic states – we define a smearing operator on a time-slice which acts in 3-space ($\vec{x}$) and color ($i$)

$$\square(\vec{x}i, \vec{y}j; t) = \sum_{n=1}^{N} \xi_n(\vec{x}i; t)\,\xi_n^\dagger(\vec{y}j; t), \qquad (6)$$

where one option is to choose $\{\xi_n\}$ to be the lowest $N$ eigenvectors of the gauge-covariant Laplacian on timeslice $t$. If all the quark fields in a correlator are smeared by application of this operator, the combination of eigenvectors and the Dirac matrix inverse, $M^{-1}$, called a 'perambulator', $\xi_n^\dagger(t')M^{-1}(t', t)\xi_m(t) \equiv \tau_{nm}(t', t)$ will appear in any propagation. Thus the basic numerical problem to be solved is inversion of the Dirac matrix on sources, $\{\xi_n\}_{n=1...N}$, which is a smaller vector space than that of the full lattice. A detailed presentation of the properties of distillation can be found in Ref. [15].

Five basic topologies of diagram appear when one Wick contracts correlators in a basis of $\bar{\psi}\boldsymbol{\Gamma}\psi$ and $\bar{\psi}\boldsymbol{\Gamma}^A\psi \cdot \bar{\psi}\boldsymbol{\Gamma}^B\psi$ operators projected into overall isospin-1 – these are shown in Fig. 4. All five topologies require light-quark perambulators between fixed $t_{\rm src}$ and varying $t$, and these are obtained by inversion of the Dirac matrix from a source on $t_{\rm src}$. While the propagation from $t_{\rm src}$ to $t_{\rm src}$ seen in topologies $A_{42}$ and $A_{44}$ does not require any extra inversions, the propagation from $t$ to $t$ seen in $A_{24}$ and $A_{44}$ requires inversion from sources at every $t$ we wish to consider. Such $\tau(t, t)$ perambulators, computed for both light and strange quarks on the $16^3$ lattice, were previously used in [13] in the computation of isoscalar correlators featuring $\bar{\psi}\boldsymbol{\Gamma}\psi$ operators. In this paper we use $\tau(t, t)$ perambulators computed on $16^3, 20^3, 24^3$ lattices for all 128 timeslices.

In order to reduce statistical noise, and make more use of each configuration, we compute correla-tors for several values of $t_{\rm src}$ and average over them, $C(t) = \frac{1}{N_{\rm src}} \sum_{t_{\rm src}} C(t + t_{\rm src}, t_{\rm src})$. Although we have inverted from every timeslice on the lattice (in order to have the $\tau(t, t)$ perambulators), covariance between neighbouring timeslices reduces the effectiveness of averaging over $t_{\rm src}$ once the sources get too close together, so in practice, given the computational cost of contracting perambulators into correlators, we average only over a handful of well-separated $t_{\rm src}$ values, typically fewer than 8 per configuration.

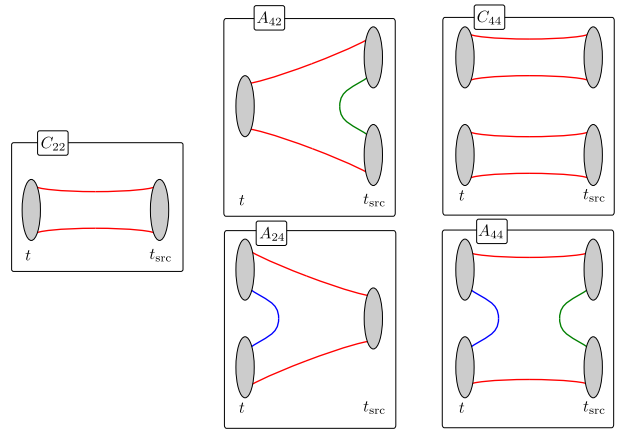

FIG. 4. A schematic of the basic Wick-contraction topologies required to compute a matrix of correlation functions in a $\{\bar{\psi}\boldsymbol{\Gamma}\psi, \bar{\psi}\boldsymbol{\Gamma}^A\psi \cdot \bar{\psi}\boldsymbol{\Gamma}^B\psi\}$ basis with isospin-1. $t_{\rm src}$ is fixed and the time dependence with respect to $t - t_{\rm src}$ is studied.

As an example of the distillation contraction for $\bar{\psi}\boldsymbol{\Gamma}\psi \cdot \bar{\psi}\boldsymbol{\Gamma}\psi$ at both source and sink, choosing the temporal origin to be at $t_{\rm src} = 0$, we have

$$\left\langle \bar{\psi}_t \square \boldsymbol{\Gamma}_t^A \square \psi_t\ \bar{\psi}_t \square \boldsymbol{\Gamma}_t^B \square \psi_t \cdot \bar{\psi}_0 \square \boldsymbol{\Gamma}_0^C \square \psi_0\ \bar{\psi}_0 \square \boldsymbol{\Gamma}_0^D \square \psi_0 \right\rangle,$$

where the 3-space ($\vec{x}$) and color ($i$) dependence of $\boldsymbol{\Gamma}(t)_{\vec{x}i, \vec{y}j}$ might include gauge-covariant spatial derivatives and/or a projection into definite momentum. Integration of the quark fields leads to terms which include the Wick contraction $A_{44}$,

$$A_{44} = \tau_{vn}(0, t)\Phi^A_{nm}(t)\tau_{mp}(t, t)\Phi^B_{pq}(t)$$
$$\times\ \tau_{qr}(t, 0)\Phi^C_{rs}(0)\tau_{su}(0, 0)\Phi^D_{uv}(0),$$

where the repeated distillation indices are summed over and where the Dirac-spin indices are suppressed. This factorised form is such that any operator construction, $\boldsymbol{\Gamma}$ (embedded in $\Phi_{nm} \equiv \xi_n^\dagger \boldsymbol{\Gamma}\xi_m$), can be considered. For example, if there is a momentum projection, a Dirac spin projection ($\Gamma^{\alpha\beta}$) and gauge-covariant derivatives ($\overleftrightarrow{D}$), then

$$\Phi^{\alpha\beta}_{nm}(\vec{p}; t)$$
$$= \Gamma^{\alpha\beta} \sum_{\vec{x}i, \vec{y}j} \xi_n^\dagger(\vec{x}i; t)\, e^{i\vec{p}\cdot\vec{x}}\left[\overleftrightarrow{D}\dots\right]_{\vec{x}i, \vec{y}j}(t)\, \xi_m(\vec{y}j; t),$$

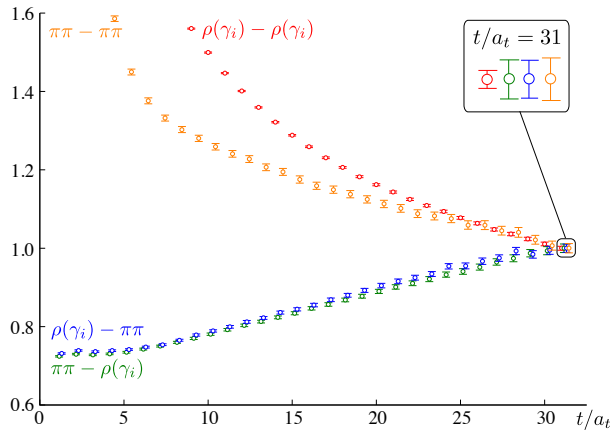

FIG. 5. $\vec{P} = [001]$, $A_1$ correlators evaluated on the $24^3$ lattice constructed from $\bar{\psi}\gamma_i\psi$ and the lowest momentum $\pi\pi$ operator $\sim \pi([001])\pi([000])$. Shown is $e^{Et}C(t)$ with $a_t E = a_t E_{\text{thr.}} \equiv \sqrt{(2a_t m_\pi)^2 + \left(\frac{2\pi}{\xi L/a_s}\right)^2}$. Normalised to all be equal to 1 at $t/a_t = 31$. The inset shows the data at $t/a_t = 31$ indicating the relative statistical noise on the correlators.

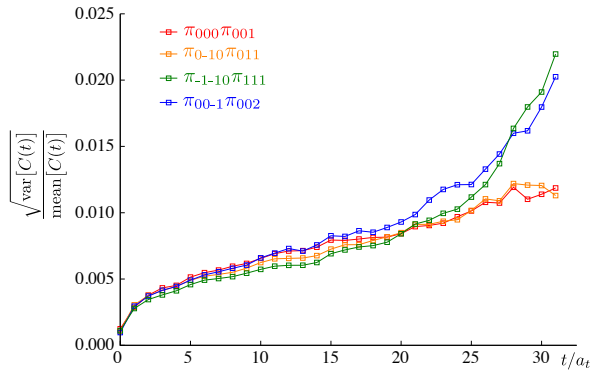

FIG. 6. $\vec{P} = [001]$, $A_1$ correlators evaluated on the $24^3$ lattice. The statistical noise, relative to signal, is shown for $\pi\pi$ diagonal correlators for a set of pion momenta.

and provided $e^{i\vec{p}\cdot\vec{x}}\left[\overleftrightarrow{D}\ldots\right]_{\vec{x}i,\vec{y}j}(t)$ is well-supported in the restricted space of vectors $\left\{\xi_n(\vec{x}i)\right\}_{n=1\ldots N}$, distillation should provide an efficient method to compute the correlator. In practice [9, 11, 15, 17] we find that the modest numbers of vectors, $N$, given in Table I, are sufficient to obtain excellent signals for hadrons with momenta $p^2 \lesssim 6\left(\frac{2\pi}{L}\right)^2$.

We demonstrate the quality of the determined correlators in Figs. 5 and 6. In Fig. 5 we present correlators, evaluated on the $24^3$ lattice in the $A_1$ irrep of $\vec{P} = [001]$, constructed using the simplest fermion bilinear operator, $\bar{\psi}\gamma_i\psi$, and the lowest momentum $\pi\pi$ operator $\sim \pi([001])\pi([000])$. We see that correlators featuring multi-pion operators are not much noisier than simple fermion bilinears and that both types of operator are likely to have a significant overlap with the ground state and differing overlaps with excited states. In Fig. 6 we show that the statistical noise on correlators does not grow rapidly with increase in the momentum of pions in $\pi\pi$ operator constructions.

## VII. FINITE-VOLUME SPECTRA

We computed correlation matrices for a range of total momentum $\vec{P}$ and irreps $\Lambda$ using the fermion-bilinear plus $\pi\pi$ operator basis described in Section V. The corresponding finite-volume spectra were determined by application of the variational method described in Section III. As an illustration, we show in Fig. 7 the 'principal correlators', $\lambda_{\mathbf{n}}(t)$, for the lowest eight levels (from a total of 22) in the case $\vec{P} = [001]$, $\Lambda = A_1$ on the $24^3$ lattice. The fits to the time-dependences of these determine the energies with high precision. The corresponding operator overlap structure was previously displayed in the first column of Fig. 1. The effective masses of the five lowest principal correlators are shown together in the bottom-right pane of Fig. 1.

In Fig. 8 we show the extracted volume dependence of the spectra for $\vec{P} = [001]$ for the irreps, $A_1$ and $E_2$. Also shown are the energy thresholds for various inelastic processes and the energies of non-interacting meson-meson states $\pi\pi$, $K\overline{K}$, $\pi\omega$. Clearly we are not observing the expected number of levels in the inelastic region and we expect that this is caused by our not using operator constructions with good overlap onto e.g. $K\overline{K}$. As mentioned in Section V, in the $E_2$ irrep, parity is not a good quantum number and contributions from $J^P = 1^+$ can appear – indeed the points shown in grey at $a_t E_{\text{cm}} \sim 0.25$ appear to have large overlap with operators characteristic of the $b_1$ meson. Since the operator basis we have used does not sample well the inelastic spectrum, we will restrict our analysis to the elastic region between $\pi\pi$ and $K\overline{K}$ thresholds.

In Fig. 9 we show the volume dependence of the spectra across $\vec{P}$ and $\Lambda$ in the elastic region. For the phase-shift analysis which follows in the next section we exclude points very close to $K\overline{K}$ threshold since we have not included $K\overline{K}$-like operators in the basis. Levels determined on the $16^3$ lattice, which potentially suffer from the largest unwanted exponential $m_\pi L$ dependence and which are often noisy, are mostly excluded from the phase-shift analysis.

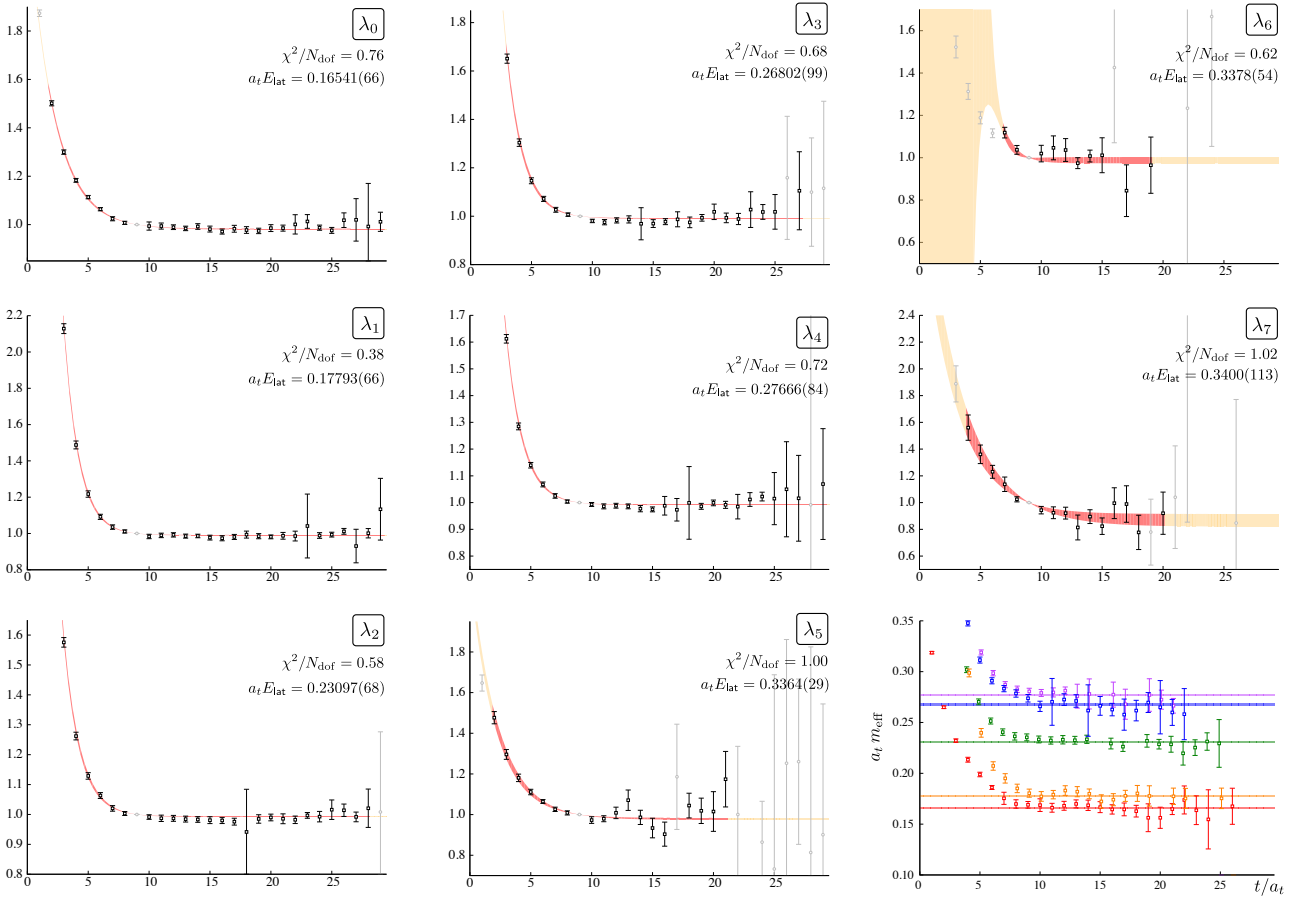

FIG. 7. Result of variational analysis of $\vec{P} = [001]$, $A_1$ correlation matrix. For the lowest eight levels, we show the "principal correlators", the eigenvalues $\lambda_{\mathrm{n}}(t)$ of the variational method, Eq. 2. Plotted as $e^{E_{\mathrm{n}}(t-t_0)}\lambda_{\mathrm{n}}(t)$ are the data and the fit according to the form $\lambda_{\mathrm{n}}(t) = (1 - A_{\mathrm{n}})\, e^{-E_{\mathrm{n}}(t-t_0)} + A_{\mathrm{n}}\, e^{-E'_{\mathrm{n}}(t-t_0)}$. Energies are those measured in the rest frame of the lattice.

The bottom-right pane shows effective masses of $\lambda_{\mathrm{n}}(t)$ for the lowest five levels. Horizontal lines indicate the fitted energies, $a_t E_{\mathrm{n}}$

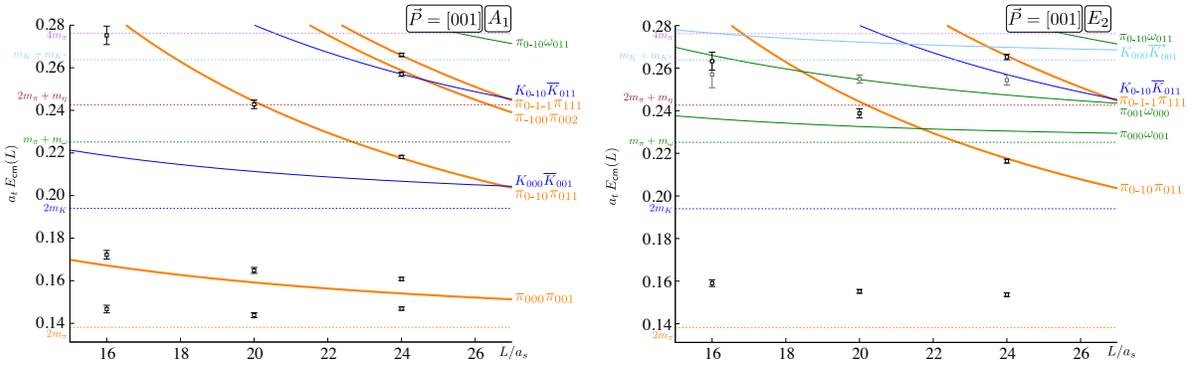

FIG. 8. Volume dependence of $\vec{P} = [001]$ spectrum in irreps $A_1$ and $E_2$. Black and gray points show the extracted spectrum. Dashed lines show energy thresholds and solid lines the non-interacting meson-pair energies.

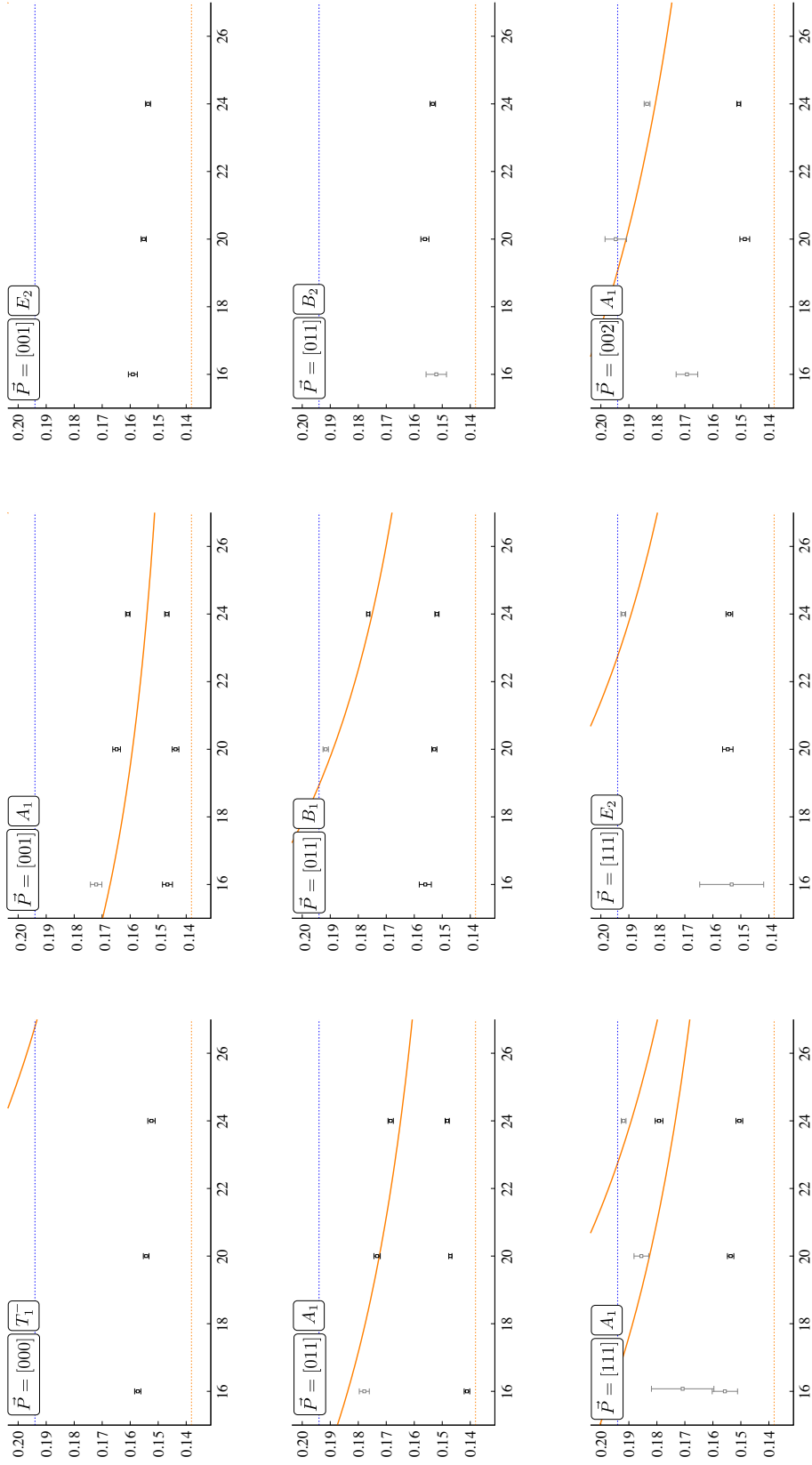

FIG. 9. Volume dependence of elastic spectra for various $\vec{P}$, $\Lambda$. Plotted is $a_t E_{\mathsf{cm}}$ versus $L/a_s$. Also shown by dashed horizontal lines are the $\pi\pi$ and $K\overline{K}$ energy thresholds. Solid curves indicate the non-interacting $\pi\pi$ energy levels. Points shown in gray are excluded from the phase-shift analysis in the following section.

## VIII. ELASTIC SCATTERING PHASE SHIFT

Our implementation of the formalism relating elastic $\pi\pi$ scattering phase-shifts to the finite-volume spectrum at rest and in flight is presented in Section VIII of [17]. Here we repeat only the essential feature, that the phase-shifts are related to the finite-volume energies by a formula which for $\pi\pi$ in $I = 1$ takes the form,

$$\det\left[\begin{pmatrix} e^{2i\delta_1(E_{cm})} & 0 & \\ 0 & e^{2i\delta_3(E_{cm})} & \cdots \\ & \vdots & \ddots \end{pmatrix} - \mathbf{U}^{\vec{P},\Lambda}\left(\left(\frac{p_{cm}L}{2\pi}\right)^2\right)\right] = 0. \tag{7}$$

The matrices here, in the space of $\ell$ contributing to the irrep, are formally infinite dimensional, with $\mathbf{U}$ being a matrix of known functions particular to the little group irrep $(\vec{P}, \Lambda)$. Which values of $\ell$ contribute to a given $(\vec{P}, \Lambda)$ is determined by the subductions given in Table III. We presume we can truncate the matrices to a finite size since at low energy we expect $\delta_{\ell>3} \ll \delta_3 \ll \delta_1$.

In Fig. 10 we present the determined phase-shifts assuming that all $\delta_{\ell\geq3}$ are negligible throughout the elastic region. In this case Eq. 7 becomes one equation in one unknown, $\delta_1(E_{cm})$, for each energy level, $E_{cm}(\vec{P}, \Lambda, \mathfrak{n})$ in the elastic region. The assumption $\delta_{\ell>1} \approx 0$ will be shown to be justified in Subsection VIII B.

### A. Resonant Parameterisations

The phase-shift in Fig. 10 shows clear resonant behavior in a small region around $a_t E_{cm} = 0.152$ which suggests we might attempt to describe $\delta_1(E_{cm})$ using a parameterisation featuring a single resonance. A popular choice is a variant of the relativistic Breit-Wigner with an energy-dependent width enforcing the $p_{cm}^{2\ell+1}$ behavior as $p_{cm} \to 0$ mandated by angular momentum conservation:

$$\tan\delta_1(E_{cm}) = \frac{E_{cm}\,\Gamma_1(E_{cm})}{m_R^2 - E_{cm}^2}$$

$$\text{where}\quad \Gamma_{\ell=1}^{BW}(E_{cm}) = \frac{g^2}{6\pi}\frac{p_{cm}^3}{E_{cm}^2}. \tag{8}$$

The parameterisation of the width is in terms of a coupling constant, $g$, which, it has been suggested, is largely independent of the quark mass [19, 33].

We choose to fit the energy levels directly within a $\chi^2$ function of the form,

$$\chi^2(\{a_i\}) = \sum_L \sum_{\substack{\vec{P}\Lambda\mathfrak{n} \\ \vec{P}'\Lambda'\mathfrak{n}'}} \left[E_{cm}(L; \vec{P}\Lambda\mathfrak{n}) - E_{cm}^{par.}(L; \vec{P}\Lambda\mathfrak{n}; \{a_i\})\right] \mathbb{C}^{-1}(L; \vec{P}\Lambda\mathfrak{n}; \vec{P}'\Lambda'\mathfrak{n}')\left[E_{cm}(L; \vec{P}'\Lambda'\mathfrak{n}') - E_{cm}^{par.}(L; \vec{P}'\Lambda'\mathfrak{n}'; \{a_i\})\right] \tag{9}$$

where $E_{cm}^{par.}(L; \vec{P}\Lambda\mathfrak{n}; \{a_i\})$ is the $\mathfrak{n}^{th}$ solution of Eq. 7 with a parameterised phase-shift depending upon parameters with values $\{a_i\}$, e.g. the resonance mass $m_R$ and the coupling, $g$. Data covariance, $\mathbb{C}$, whose off-diagonal elements between energies evaluated on the same ensemble can be non-zero, is estimated using jackknife.

Fitting energy levels with $a_t E_{cm} < 0.18$ to the Breit-Wigner form in Eq. 8, we obtain the following parameter values, errors and correlation,

$$a_t m_R = 0.15241(33)(10) \qquad \begin{bmatrix} 1 & -0.09 \\ & 1 \end{bmatrix}$$
$$g = 4.83(13)(2)$$

$$\chi^2/N_{dof} = \frac{48.9}{29-2} = 1.81,$$

where the first error is statistical and the second error reflects variation of $m_\pi$ and $\xi$ within their respective uncertainties (see [17]). There is observed to be very little correlation between the mass and the coupling constant. The energy dependence of the resulting $\delta_1(E_{cm})$ is shown by the blue curve in Fig. 11.

A criticism of the parameterisation in Eq. 8 is that the $p_{cm}^3$ form, introduced to give the right threshold behaviour, continues to grow at large energies in an unrealistic way. As seen in Fig. 11, the phase-shift approaches 180° very slowly, allowing the effect of the resonance to be felt many half-widths above the mass. Generally one expects the $p_{cm}^3$ behaviour to be damped at larger energies, and one physically motivated approach is to appeal to the idea that the resonance has a finite spatial size, and introduce 'barrier factors'. One implementation, commonly used in experimental partial wave analysis, follows von Hippel & Quigg [34], by considering the interaction giving rise to the resonance to have a sharp size, $R$. The resulting scattering wavefunctions are outgoing spherical Hankel functions which give rise to polynomial damping factors. In the case $\ell = 1$ this leads to the following modification of the energy-dependent width:

$$\Gamma_{\ell=1}^{HQ}(E_{cm}) = \frac{g^2}{6\pi}\frac{p_{cm}^3}{E_{cm}^2}\frac{1 + (p_R R)^2}{1 + (p_{cm}R)^2}, \tag{10}$$

where $p_R$ is the cm scattering momentum at the resonance mass, $p_R = \frac{1}{2}\sqrt{m_R^2 - 4m_\pi^2}$. The barrier factor is normalised such that it does not affect the width evaluated at the resonance mass. We fit the same data as above to obtain

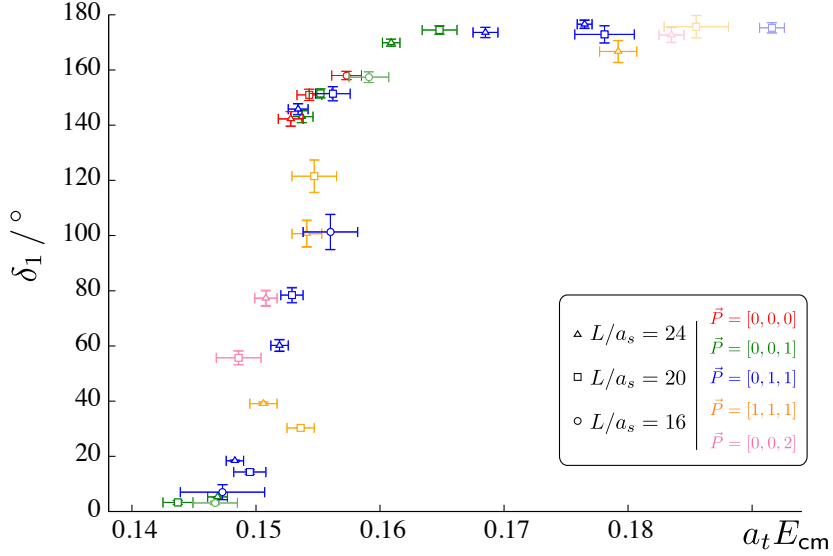

FIG. 10. $P$-wave $\pi\pi$ elastic scattering phase-shift, $\delta_1(E_{cm})$, determined from solution of Eq. 7 applied to the finite-volume spectra shown in Fig. 9 under the assumption that $\delta_{\ell>1} = 0$. Energy region plotted is from $\pi\pi$ threshold to $K\overline{K}$ threshold.

$$a_t m_R = 0.15226(34)(11)$$
$$g = 5.06(15)(2)$$
$$R/a_t = 16.6(52)(17)$$

$$\begin{bmatrix} 1 & -0.14 & -0.09 \\ & 1 & 0.32 \\ & & 1 \end{bmatrix}$$

$$\chi^2/N_{dof} = \frac{43.6}{29-3} = 1.68,$$

which shows a slightly improved quality of fit, although there is clearly some correlation between the coupling $g$ and the range $R$. The range expressed in physical units $R = \frac{R}{a_t} \frac{a_t m_\Omega}{m_\Omega^{phys}} \approx 0.6 \pm 0.2\,\text{fm}$ would seem to be reasonable on the usual hadronic scale. The resulting energy dependence is shown by the red curve in Fig. 11 where it is seen to approach 180° more rapidly than the simple Breit-Wigner.

The particular form of the damping function is a model-dependent choice and we can explore the sensitivity by trying other parameterisations. For example a gaussian form (previously considered in a quark model study [35]),

$$\Gamma_{\ell=1}^{gau}(E_{cm}) = \frac{g^2}{6\pi} \frac{p_{cm}^3}{E_{cm}^2} \frac{e^{-p_{cm}^2/6\beta^2}}{e^{-p_R^2/6\beta^2}}. \tag{11}$$

Fitting the same dataset we obtain

$$a_t m_R = 0.15224(34)(14)$$
$$g = 5.08(17)(3)$$
$$a_t\beta = 0.029(7)(3)$$

$$\begin{bmatrix} 1 & -0.18 & 0.16 \\ & 1 & -0.47 \\ & & 1 \end{bmatrix}$$

$$\chi^2/N_{dof} = \frac{43.5}{29-3} = 1.67,$$

indicating that the particular functional form of the damping appears to be relatively unimportant. In physical units, $\beta = a_t\beta \cdot \frac{m_\Omega^{phys}}{a_t m_\Omega} \approx 160(40)\,\text{MeV}$. The energy dependence is shown by an orange curve in Fig. 11 that lies almost exactly on the red curve already described.

Another parameterisation that has been used to fit experimental phase-shift data is provided by Peláez and Yndráin (see Ref. [36] and their subsequent papers),

$$\cot\delta_1(E_{cm}) = \frac{E_{cm}}{2p_{cm}^3}(m_R^2 - E_{cm}^2)$$
$$\times \left[\frac{2m_\pi^3}{m_R^2 E_{cm}} + B_0 + B_1\frac{E_{cm} - \sqrt{s_0 - E_{cm}^2}}{E_{cm} + \sqrt{s_0 - E_{cm}^2}}\right],$$

which, while it appears cosmetically to be very different to a Breit-Wigner, in fact has an energy dependence which is rather similar, with the three parameters $m_R, B_0, B_1$ able to conspire to provide damping. The additional parameter, $s_0$, is not allowed to float, and following the proposers' suggestion is set to $2m_\pi + m_\rho$, as determined on this lattice, $a_t\sqrt{s_0} = 0.29$. Fitting yields

$$a_t m_R = 0.15227(34)(12)$$
$$B_0 = 2.71(77)(21)$$
$$B_1 = 6.0(33)(9)$$

$$\begin{bmatrix} 1 & -0.06 & -0.05 \\ & 1 & 0.99 \\ & & 1 \end{bmatrix}$$

$$\chi^2/N_{dof} = \frac{43.7}{29-3} = 1.68,$$

a reasonable description of the data. The extremely high degree of correlation between $B_0$ and $B_1$ suggests that they may not be the most natural way to parameterise this amplitude. The energy dependence is plotted in Fig. 11 using a green curve that lies almost exactly under the orange and red curves already plotted.

We have presented the data and fits in units of the inverse temporal lattice spacing thus far to avoid ambiguities with how one sets the lattice scale. If we choose our usual scale setting procedure where $a_t = \frac{a_t m_\Omega}{m_\Omega^{phys}}$ using the $\Omega$ baryon mass determined on these lattices ($a_t m_\Omega = 0.2951$) and the physical $\Omega$ baryon mass $m_\Omega^{phys} = 1672$ MeV, then the simple Breit-Wigner fit corresponds to $m_R = 863.5(19)(6)$ MeV and $\Gamma_R = 10.1(6)(1)$ MeV. As expected in a calculation with heavier than physical mass light quarks, the resonance mass is somewhat larger than the physical $\rho$ mass. The small width is explained by the much-reduced phase-space for decay of an 864 MeV resonance into two pions of mass 391 MeV compared with the physical kinematics. We observe from the rather similar $\chi^2/N_{dof}$ that the data do not clearly distinguish between the various parameterisations which vary only in the tails of the resonance. This may be due to the very narrow nature of the resonance with the small phase space for decay.

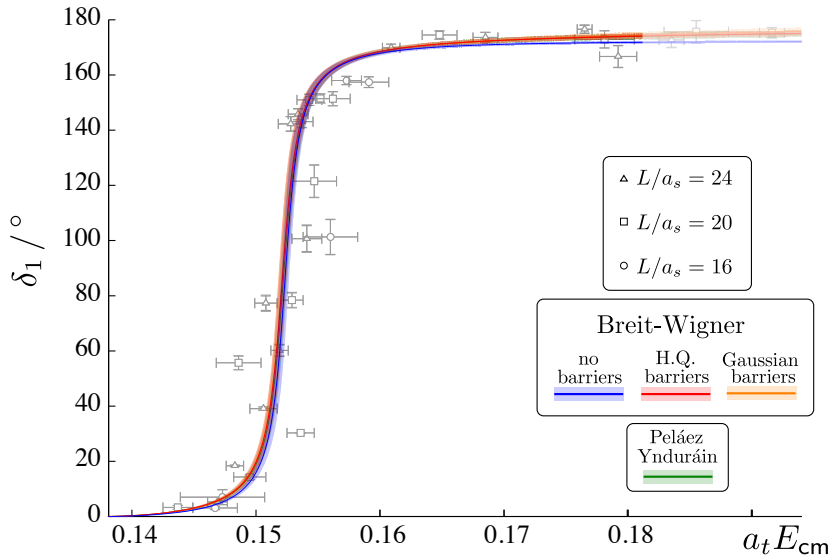

FIG. 11. $P$-wave $\pi\pi$ elastic scattering phase-shift, $\delta_1(E_{cm})$, as determined by describing the finite-volume spectra by resonant parameterisations as described in the text. The barrier factor variations and the Peláez & Ynduráin fit all lie on top of each other. Also shown as gray points are the data previously presented in Fig. 10. Energy region plotted is from $\pi\pi$ threshold to $K\overline{K}$ threshold.

## B. Role of higher partial-waves

From Eq. 7 and Table III it is apparent that in principle many partial waves contribute to the determination of the finite-volume spectrum in each irrep, in particular when the system is in-flight. The next lowest $\ell$ that can contribute in $\pi\pi$ $I = 1$ scattering is $\ell = 3$ which is leading in e.g. irreps ($\vec{P} = [001]$, $B_1$) and ($\vec{P} = [001]$, $B_2$). For the lattice volumes we consider, the lowest energy level in these irreps is always above the elastic region, and as such we cannot apply Eq. 7 without concern about neglecting other open channels (in this case $K\overline{K}$). If we assume that there is zero coupling into $K\overline{K}$ and proceed in a cavalier manner with application of Eq. 7 we obtain points at energies only slightly above the $K\overline{K}$ threshold that have $\delta_3$ compatible with zero (roughly $(-1 \pm 1)^\circ$).

One way to obtain estimates of $\delta_3$ in the elastic regime is to consider a number of approximately degenerate energy levels coming from different irreps. By writing a version of Eq. 7 for each one we can approximately solve that coupled set of equations for $\delta_1, \delta_3$ at the relevant energy. This approach was described in some detail for $\pi\pi$ $I = 2$ scattering in [17]. An example set of levels is ($[000], T_1^-, \mathfrak{n} = 0$), ($[001], E_2, \mathfrak{n} = 0$) and ($[011], B_2, \mathfrak{n} = 0$) which on the $24^3$ lattice all have an energy $a_t E_{cm} \approx 0.153(1)$. Solving the coupled system of equations we find $\delta_1 = 145.7(22)^\circ$ and $\delta_3 = -0.048(55)^\circ$. The same set of levels on the $20^3$ lattice have $a_t E_{cm} \approx 0.155(1)$ and give $\delta_1 = 151.1(30)^\circ$ and $\delta_3 = +0.002(24)^\circ$.

We also tried parameterised fits to all data points, as in the previous section but including a scattering length parameterisation for the $\ell = 3$ wave, $p_{cm}^7 \cot\delta_3 = 1/a_3$, as well as a resonant parameterisation of $\delta_1(E_{cm})$. The fits were essentially the same quality (in $\chi^2/N_{dof}$) and gave $a_3 = -3.4(33)(6) \times 10^6 \cdot a_t^7$ with a negligible change in the $\ell = 1$ Breit-Wigner parameters. This parameterisation gives $\delta_3 = -1.3(13)^\circ$ at the $K\overline{K}$ threshold.

In summary, the lattice data require no non-zero value of $\delta_3$ throughout the elastic region and our analysis in the previous section based upon $\delta_3 = 0$ is justified.

These observations (at $m_\pi \sim 400$ MeV) are in accord with experimental expectations (at the physical pion mass). In the $\pi\pi$ partial wave analysis of Estabrooks and Martin [37], no $F$-wave amplitude was required in the elastic region to describe $\pi^- p \to \pi^- \pi^+ n$ data. Above inelastic threshold, the $F$-wave amplitude between $E_{\text{cm}} = 1.5$ and $1.9$ GeV was well described by a Breit-Wigner form with barrier factors. Subsequent experiments have determined the resonance parameters of the $\rho_3(1690)$ with greater precision [38]. The tail of this resonance in the elastic region (which must vanish like $p_{\text{cm}}^7$) is such that $\delta_3 \sim +1.8°$ at the $K\overline{K}$ threshold. The energy-dependent phase-shift analysis of $\pi^+ p \to \pi^+ \pi^- \Delta^{++}$ data by Protopoescu et al. [39] suggests a $\delta_3$ which is small and negative below $K\overline{K}$ threshold, with the largest deviation from zero $\sim -1.5(5)°$.

## IX. SUMMARY

In summary, we have used lattice field theory methods to compute part of the discrete energy spectrum of QCD in finite boxes. Through the known connection of the discrete spectrum to elastic scattering amplitudes, we have mapped out the energy dependence of the $\pi\pi$ isospin-1 $P$-wave scattering phase shift in unprecedented detail up to the $K\overline{K}$ threshold. The data, presented in Fig. 10, unambiguously show the rapidly-rising form expected in the presence of an elastic resonance. As observed in Fig. 11, the phase shift can be well described by a single $\rho$ resonance. While in principle the discrete spectrum can be sensitive to scattering in higher-partial waves, we find that no non-zero value of $F$-wave (or higher) phase-shift is required in the elastic region.

The $P$-wave scattering phase-shift is summarised in Fig. 12, where the scale in all dimensionful quantities is set using $a_t = \frac{a_t m_\Omega}{m_\Omega^{\text{phys}}}$. A simple Breit-Wigner fit describes the data reasonably well in terms of a single narrow resonance. We note that the extracted coupling $g$ is compatible with other lattice determinations [18–22].

This work represents a successful application in a resonant channel of the methodology for lattice computations of scattering phase shifts presented in Ref. [17], where it was initially applied to a non-resonant channel. *Distillation* enabled us to efficiently compute correlators with large bases of carefully constructed fermion-bilinear and $\pi\pi$-like interpolating operators, in various irreducible representations of the relevant symmetry group and for a range of center-of-mass momenta, with high statistical precision. In addition, the inclusion of the necessary quark annihilation contributions is rendered straightforward. Variational analyses of the resulting correlators gave a large number of finite-volume energy levels which, in turn, allowed us to determine an unparalleled number of points on the phase shift curve.

We plan to compute scattering in other channels using the same technology, including situations where experimental and phenomenological understanding is incomplete and where we cannot assume simple parameterisations will describe the data. In these cases, mapping out the phase shift in detail will be more important than in the simple case of elastic $\pi\pi$ scattering in isospin-1 that we have considered here, where a single vector resonance is expected to dominate the scattering amplitude. Determining the energy dependence of the scattering amplitude in detail will inform the parameterisations. Future work to explore the hadron resonance spectrum will need to consider inelastic scattering - we can reliably extract the finite-volume spectrum above inelastic thresholds by including the appropriate meson-meson-like operators (e.g. $K\overline{K} \ldots$), following the general methodology developed in Ref. [17].

## ACKNOWLEDGMENTS

We thank our colleagues within the Hadron Spectrum Collaboration. Particular thanks to C. Shultz for his updates to our variational fitting code. `Chroma` [40] and `QUDA` [41, 42] were used to perform this work on clusters at Jefferson Laboratory under the USQCD Initiative and the LQCD ARRA project. Gauge configurations were generated using resources awarded from the U.S. Department of Energy INCITE program at Oak Ridge National Lab, the NSF Teragrid at the Texas Advanced Computer Center and the Pittsburgh Supercomputer Center, as well as at Jefferson Lab. RGE and JJD acknowledge support from U.S. Department of Energy contract DE-AC05-06OR23177, under which Jefferson Science Associates, LLC, manages and operates Jefferson Laboratory. JJD also acknowledges the support of the Jeffress Memorial Fund and the U.S. Department of Energy Early Career award contract DE-SC0006765. CET acknowledges support from a Marie Curie International Incoming Fellowship, PIIF-GA-2010-273320, within the 7th European Community Framework Programme. RGE thanks the Galileo Galilei Institute for Theoretical Physics for the hospitality and the INFN for partial support during the completion of this work.

## Appendix A: Finite-volume phase-shift relations for $\delta_{\ell > 1} = 0$

A general procedure for generating the little-group irrep matrix-elements $\mathcal{M}_{\ell n; \ell' n'}^{(\vec{P}, \Lambda, \mu)}$, which are used to construct the matrix $\mathbf{U}^{\vec{P}, \Lambda}$ in Eqn 7, is provided in [17], along with the required subductions for all $\ell \leq 4$. In the case of equal-mass pseudoscalar-pseudoscalar scattering (as is relevant in this $\pi\pi$ case), and assuming $\delta_{\ell > 1} = 0$, we can reduce Eqn 7 to the following simple forms for various $\vec{P} = [n_x n_y n_z]$ and irreps,

$$[00n] \quad A_1 : \quad \cot\delta_1(E_{\mathsf{cm}}) = \frac{1}{\gamma\,\pi^{3/2}\,q}\left[Z_{0,0}^{[00n]}(q^2) + \frac{2}{\sqrt{5}}\frac{1}{q^2}Z_{2,0}^{[00n]}(q^2)\right]$$

$$[00n] \quad E_2 : \quad \cot\delta_1(E_{\mathsf{cm}}) = \frac{1}{\gamma\,\pi^{3/2}\,q}\left[Z_{0,0}^{[00n]}(q^2) - \frac{1}{\sqrt{5}}\frac{1}{q^2}Z_{2,0}^{[00n]}(q^2)\right]$$

$$[0nn] \quad A_1 : \quad \cot\delta_1(E_{\mathsf{cm}}) = \frac{1}{\gamma\,\pi^{3/2}\,q}\left[Z_{0,0}^{[0nn]}(q^2) + \frac{1}{2\sqrt{5}}\frac{1}{q^2}Z_{2,0}^{[0nn]}(q^2) + i\sqrt{\frac{6}{5}}\frac{1}{q^2}Z_{2,1}^{[0nn]}(q^2) - \sqrt{\frac{3}{10}}\frac{1}{q^2}Z_{2,2}^{[0nn]}(q^2)\right]$$

$$[0nn] \quad B_1 : \quad \cot\delta_1(E_{\mathsf{cm}}) = \frac{1}{\gamma\,\pi^{3/2}\,q}\left[Z_{0,0}^{[0nn]}(q^2) + \frac{1}{2\sqrt{5}}\frac{1}{q^2}Z_{2,0}^{[0nn]}(q^2) - i\sqrt{\frac{6}{5}}\frac{1}{q^2}Z_{2,1}^{[0nn]}(q^2) - \sqrt{\frac{3}{10}}\frac{1}{q^2}Z_{2,2}^{[0nn]}(q^2)\right]$$

$$[0nn] \quad B_2 : \quad \cot\delta_1(E_{\mathsf{cm}}) = \frac{1}{\gamma\,\pi^{3/2}\,q}\left[Z_{0,0}^{[0nn]}(q^2) - \frac{1}{\sqrt{5}}\frac{1}{q^2}Z_{2,0}^{[0nn]}(q^2) + \sqrt{\frac{6}{5}}\frac{1}{q^2}Z_{2,2}^{[0nn]}(q^2)\right]$$

$$[nnn] \quad A_1 : \quad \cot\delta_1(E_{\mathsf{cm}}) = \frac{1}{\gamma\,\pi^{3/2}\,q}\left[Z_{0,0}^{[nnn]}(q^2) - i\sqrt{\frac{8}{15}}\frac{1}{q^2}Z_{2,2}^{[nnn]}(q^2) - \sqrt{\frac{8}{15}}\frac{1}{q^2}\mathrm{Re}\left[Z_{2,1}^{[nnn]}(q^2)\right] - \sqrt{\frac{8}{15}}\frac{1}{q^2}\mathrm{Im}\left[Z_{2,1}^{[nnn]}(q^2)\right]\right]$$

$$[nnn] \quad E_2 : \quad \cot\delta_1(E_{\mathsf{cm}}) = \frac{1}{\gamma\,\pi^{3/2}\,q}\left[Z_{0,0}^{[nnn]}(q^2) + i\sqrt{\frac{6}{5}}\frac{1}{q^2}Z_{2,2}^{[nnn]}(q^2)\right]$$

where $Z_{\ell,m}^{\vec{P}}(q^2)$ is the generalized zeta function [4] with argument $q^2 = \left(\frac{p_{cm}L}{2\pi}\right)^2$.

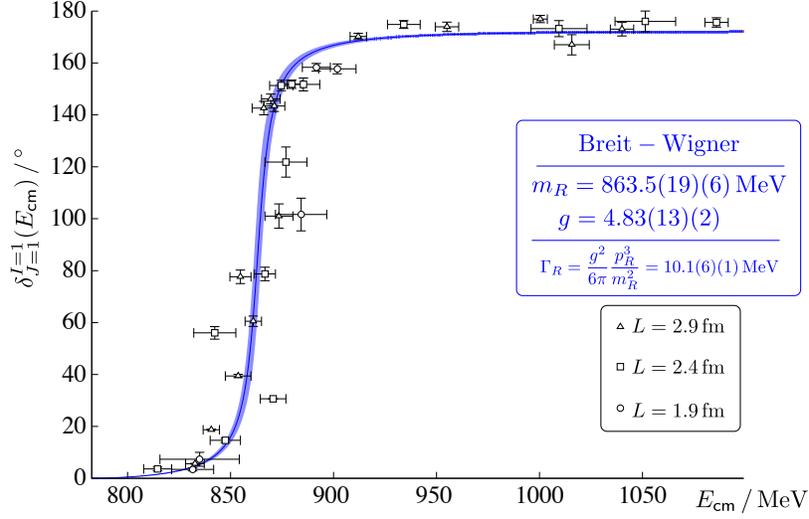

FIG. 12. Isospin-1, $P$-wave $\pi\pi$ elastic scattering phase shift and Breit-Wigner parameterisation for $m_\pi = 391\,\mathrm{MeV}$. Energy region plotted is from $\pi\pi$ threshold to $K\overline{K}$ threshold.

[1] L. Maiani and M. Testa, Phys.Lett. **B245**, 585 (1990).

[2] M. Luscher and U. Wolff, Nucl. Phys. **B339**, 222 (1990).

[3] M. Luscher, Nucl. Phys. **B364**, 237 (1991).

[4] K. Rummukainen and S. A. Gottlieb, Nucl.Phys. **B450**,

397 (1995), arXiv:hep-lat/9503028 [hep-lat].

[5] C. Kim, C. Sachrajda, and S. R. Sharpe, Nucl.Phys. **B727**, 218 (2005), arXiv:hep-lat/0507006 [hep-lat].

[6] N. H. Christ, C. Kim, and T. Yamazaki, Phys.Rev. **D72**, 114506 (2005), arXiv:hep-lat/0507009 [hep-lat].

[7] Z. Fodor and C. Hoelbling, Rev.Mod.Phys. **84**, 449 (2012), arXiv:1203.4789 [hep-lat].

[8] J. J. Dudek, R. G. Edwards, M. J. Peardon, D. G. Richards, and C. E. Thomas, Phys. Rev. Lett. **103**, 262001 (2009), arXiv:0909.0200 [hep-ph].

[9] J. J. Dudek *et al.*, Phys. Rev. **D82**, 034508 (2010), arXiv:1004.4930 [hep-ph].

[10] R. G. Edwards, J. J. Dudek, D. G. Richards, and S. J. Wallace, Phys.Rev. **D84**, 074508 (2011), arXiv:1104.5152 [hep-ph].

[11] C. E. Thomas, R. G. Edwards, and J. J. Dudek, Phys.Rev. **D85**, 014507 (2012), arXiv:1107.1930 [hep-lat].

[12] J. J. Dudek and R. G. Edwards, Phys.Rev. **D85**, 054016 (2012), arXiv:1201.2349 [hep-ph].

[13] J. J. Dudek, R. G. Edwards, B. Joo, M. J. Peardon, D. G. Richards, *et al.*, Phys.Rev. **D83**, 111502 (2011), arXiv:1102.4299 [hep-lat].

[14] L. Liu *et al.* (for the Hadron Spectrum Collaboration), JHEP **1207**, 126 (2012), arXiv:1204.5425 [hep-ph].

[15] M. Peardon *et al.* (Hadron Spectrum), Phys. Rev. **D80**, 054506 (2009), arXiv:0905.2160 [hep-lat].

[16] J. J. Dudek, R. G. Edwards, M. J. Peardon, D. G. Richards, and C. E. Thomas, Phys.Rev. **D83**, 071504 (2011), arXiv:1011.6352 [hep-ph].

[17] J. J. Dudek, R. G. Edwards, and C. E. Thomas, Phys.Rev. **D86**, 034031 (2012), arXiv:1203.6041 [hep-ph].

[18] S. Aoki *et al.* (CP-PACS Collaboration), Phys.Rev. **D76**, 094506 (2007), arXiv:0708.3705 [hep-lat].

[19] X. Feng, K. Jansen, and D. B. Renner, Phys.Rev. **D83**, 094505 (2011), arXiv:1011.5288 [hep-lat].

[20] C. Lang, D. Mohler, S. Prelovsek, and M. Vidmar, Phys.Rev. **D84**, 054503 (2011), arXiv:1105.5636 [hep-lat].

[21] S. Aoki *et al.* (CS Collaboration), Phys.Rev. **D84**, 094505 (2011), arXiv:1106.5365 [hep-lat].

[22] C. Pelissier and A. Alexandru, (2012), arXiv:1211.0092 [hep-lat].

[23] M. Gockeler, R. Horsley, M. Lage, U.-G. Meissner, P. Rakow, *et al.*, (2012), arXiv:1206.4141 [hep-lat].

[24] M. Doring, U. Meissner, E. Oset, and A. Rusetsky, Eur.Phys.J. **A48**, 114 (2012), arXiv:1205.4838 [hep-lat].

[25] L. Leskovec and S. Prelovsek, Phys.Rev. **D85**, 114507 (2012), arXiv:1202.2145 [hep-lat].

[26] R. G. Edwards, B. Joo, and H.-W. Lin, Phys. Rev. **D78**, 054501 (2008), arXiv:0803.3960 [hep-lat].

[27] H.-W. Lin *et al.* (Hadron Spectrum), Phys. Rev. **D79**, 034502 (2009), arXiv:0810.3588 [hep-lat].

[28] S. Beane *et al.* (NPLQCD Collaboration), Phys.Rev. **D85**, 034505 (2012), arXiv:1107.5023 [hep-lat].

[29] C. Michael, Nucl. Phys. **B259**, 58 (1985).

[30] B. Blossier, M. Della Morte, G. von Hippel, T. Mendes, and R. Sommer, JHEP **04**, 094 (2009), arXiv:0902.1265 [hep-lat].

[31] J. J. Dudek, R. G. Edwards, N. Mathur, and D. G. Richards, Phys. Rev. **D77**, 034501 (2008), arXiv:0707.4162 [hep-lat].

[32] D. C. Moore and G. T. Fleming, Phys. Rev. **D73**, 014504 (2006), arXiv:hep-lat/0507018.

[33] J. Pelaez and G. Rios, Phys.Rev. **D82**, 114002 (2010), arXiv:1003.1918 [hep-ph].

[34] F. Von Hippel and C. Quigg, Phys.Rev. **D5**, 624 (1972).

[35] Z. Li, M. Guidry, T. Barnes, and E. Swanson, (1994), arXiv:hep-ph/9401326 [hep-ph].

[36] J. Pelaez and F. Yndurain, Phys.Rev. **D71**, 074016 (2005), arXiv:hep-ph/0411334 [hep-ph].

[37] P. Estabrooks and A. D. Martin, Nucl.Phys. **B95**, 322 (1975).

[38] J. Beringer *et al.* (Particle Data Group), Phys.Rev. **D86**, 010001 (2012).

[39] S. Protopopescu, M. Alston-Garnjost, A. Barbaro-Galtieri, S. M. Flatte, J. Friedman, *et al.*, Phys.Rev. **D7**, 1279 (1973).

[40] R. G. Edwards and B. Joo, Nucl. Phys. B. Proc. Suppl. **140**, 832 (2005).

[41] M. A. Clark *et al.*, Comput. Phys. Commun. **181**, 1517 (2010), arXiv:0911.3191 [hep-lat].

[42] R. Babich, M. A. Clark, and B. Joo, ACM/IEEE Int. Conf. High Performance Computing, Networking, Storage and Analysis, New Orleans (2010), 10.1109/SC.2010.40, arXiv:1011.0024 [hep-lat].